\documentclass[preprint2]{aastex62}
\usepackage[bottom]{footmisc}
\usepackage{amsmath}

\newcommand{\KICPChicago}{Kavli Institute for Cosmological Physics, University of Chicago, 5640 S. Ellis Ave., Chicago, IL 60637, USA}
\newcommand{\AAUChicago}{Department of Astronomy and Astrophysics, University of Chicago, 5640 S. Ellis Ave., Chicago, IL 60637, USA}
\newcommand{\Davis}{Department of Physics, University of California, Davis, CA 95616, USA}
\newcommand{\FNAL}{Fermi National Accelerator Laboratory, MS209, P.O. Box 500, Batavia, IL 60510, USA}
\newcommand{\ArgonneHEP}{High Energy Physics Division, Argonne National Laboratory, Argonne, IL 60439, USA }
\newcommand{\PhysicsUChicago}{Department of Physics, University of Chicago, Chicago, IL 60637, USA }
\newcommand{\EFIChicago}{Enrico Fermi Institute, University of Chicago, Chicago, IL 60637, USA }
\newcommand{\SLAC}{SLAC National Accelerator Laboratory, 2575 Sand Hill Road, Menlo Park, CA 94025, USA}
\newcommand{\McGill}{Department of Physics and McGill Space Institute, McGill University, Montreal, Quebec H3A 2T8, Canada}
\newcommand{\Caltech}{California Institute of Technology, Pasadena, CA 91125, USA}
\newcommand{\Berkeley}{Department of Physics, University of California, Berkeley, CA 94720, USA }
\newcommand{\Cifar}{Canadian Institute for Advanced Research, CIFAR Program in Cosmology and Gravity, Toronto, ON, M5G 1Z8, Canada}
\newcommand{\Colorado}{Center for Astrophysics and Space Astronomy, Department of Astrophysical and Planetary Sciences, University of Colorado, Boulder, CO 80309, USA }
\newcommand{\ESO}{European Southern Observatory, Karl-Schwarzschild-Stra{\ss}e 2, 85748 Garching, Germany}
\newcommand{\Colphys}{Department of Physics, University of Colorado, Boulder, CO 80309, USA}
\newcommand{\Illast}{Astronomy Department, University of Illinois at Urbana-Champaign, 1002 W. Green Street, Urbana, IL 61801, USA}
\newcommand{\Illphys}{Department of Physics, University of Illinois Urbana-Champaign, 1110 W. Green Street, Urbana, IL 61801, USA}
\newcommand{\UChicago}{University of Chicago, Chicago, IL 60637, USA}
\newcommand{\KIPAC}{Kavli Institute for Particle Astrophysics and Cosmology, Stanford University, 452 Lomita Mall, Stanford, CA 94305, USA}
\newcommand{\LBNL}{Physics Division, Lawrence Berkeley National Laboratory, Berkeley, CA 94720, USA }
\newcommand{\Arizona}{Steward Observatory, University of Arizona, 933 North Cherry Avenue, Tucson, AZ 85721, USA}
\newcommand{\Michigan}{Department of Physics, University of Michigan, Ann  Arbor, MI 48109, USA}
\newcommand{\Munich}{Faculty of Physics, Ludwig-Maximilians-Universit\"{a}t, 81679 M\"{u}nchen, Germany}
\newcommand{\ExcellenceCluster}{Excellence Cluster Universe, 85748 Garching, Germany}
\newcommand{\MPE}{Max-Planck-Institut f\"{u}r extraterrestrische Physik, 85748 Garching, Germany}
\newcommand{\Dunlap}{Dunlap Institute for Astronomy \& Astrophysics, University of Toronto, 50 St George St, Toronto, ON, M5S 3H4, Canada}
\newcommand{\Minnesota}{Department of Physics, University of Minnesota, Minneapolis, MN 55455, USA }
\newcommand{\Melbourne}{School of Physics, University of Melbourne, Parkville, VIC 3010, Australia}
\newcommand{\CaseWestern}{Physics Department, Case Western Reserve University,Cleveland, OH 44106, USA }
\newcommand{\ArtInstChicago}{Liberal Arts Department, School of the Art Institute of Chicago, Chicago, IL 60603, USA }
\newcommand{\JPL}{Jet Propulsion Laboratory, California Institute of Technology, Pasadena, CA 91109, USA}
\newcommand{\CfA}{Harvard-Smithsonian Center for Astrophysics, Cambridge, MA 02138, USA }
\newcommand{\Stanford}{Deptartment of Physics, Stanford University, 382 Via Pueblo Mall, Stanford, CA 94305, USA}
\newcommand{\UToronto}{Department of Astronomy \& Astrophysics, University of Toronto, 50 St George St, Toronto, ON, M5S 3H4, Canada}

\DeclareGraphicsExtensions{.jpg,.pdf,.png,.eps,.ps}
\graphicspath{{FIGURES/}}

\newcommand{\beq}{\begin{equation}}
\newcommand{\eeq}{\end{equation}}
\newcommand{\simleq}{{\raise.0ex\hbox{$\mathchar"013C$}\mkern-14mu \lower1.2ex\hbox{$\mathchar"0218$}}}
\newcommand{\simgeq}{{\raise.0ex\hbox{$\mathchar"013E$}\mkern-14mu \lower1.2ex\hbox{$\mathchar"0218$}}}
\newcommand{\tcmb}{\ensuremath{T_\mathrm{CMB}}}

\newcommand{\planck}{{\sl Planck}}

\newcommand{\nside}{\ensuremath{N_\mathrm{side}}}

\newcommand{\dee}{\mathrm{d}} 
\newcommand{\mrm}{\mathrm} 

\newcommand{\hpx}{HEALPix}

\begin{document}


\title{Maps of the Southern Millimeter-wave Sky from Combined $2500\ {\rm \deg}^{2}$ SPT-SZ and \textit{Planck} Temperature Data}


\author{R. Chown}
\affiliation{\McGill}
\affiliation{Department of Physics and Astronomy, McMaster University, 1280 Main St. W., Hamilton, ON L8S 4L8, Canada}

\author{Y. Omori}
\affiliation{\KIPAC}
\affiliation{\Stanford}
\affiliation{\McGill}

%


\author{K.~Aylor}
\affiliation{\Davis}


\author{B.~A.~Benson}
\affiliation{\FNAL}
\affiliation{\KICPChicago}
\affiliation{\AAUChicago}

\author{L.~E.~Bleem}
\affiliation{\ArgonneHEP}
\affiliation{\KICPChicago}

\author{J.~E.~Carlstrom}
\affiliation{\KICPChicago}
\affiliation{\PhysicsUChicago}
\affiliation{\ArgonneHEP}
\affiliation{\AAUChicago}
\affiliation{\EFIChicago}

\author{C.~L.~Chang}
\affiliation{\ArgonneHEP}
\affiliation{\KICPChicago}
\affiliation{\AAUChicago}

\author{H-M.~Cho}
\affiliation{\SLAC}

\author{T. Crawford}
\affiliation{\KICPChicago}
\affiliation{\AAUChicago}

\author{A.~T.~Crites}
\affiliation{\KICPChicago}
\affiliation{\AAUChicago}
\affiliation{\Caltech}

\author{T.~de~Haan}
\affiliation{\McGill}
\affiliation{\Berkeley}

\author{M.~A.~Dobbs}
\affiliation{\McGill}
\affiliation{\Cifar}

\author{W.~B.~Everett}
\affiliation{\Colorado}


\author{E.~M.~George}
\affiliation{\Berkeley}
\affiliation{\ESO}

\author{J.~W.~Henning}
\affiliation{\ArgonneHEP}
\affiliation{\KICPChicago}

\author{N.~W.~Halverson}
\affiliation{\Colorado}
\affiliation{\Colphys}

\author{N.~L.~Harrington}
\affiliation{\Berkeley}

\author{G. Holder}
\affiliation{\Illast}
\affiliation{\Illphys}
\affiliation{\McGill}

\author{W.~L.~Holzapfel}
\affiliation{\Berkeley}

\author{Z.~Hou}
\affiliation{\KICPChicago}
\affiliation{\AAUChicago}

\author{J.~D.~Hrubes}
\affiliation{\UChicago}


\author{L.~Knox}
\affiliation{\Davis}

\author{A.~T.~Lee}
\affiliation{\Berkeley}
\affiliation{\LBNL}


\author{D.~Luong-Van}
\affiliation{\UChicago}

\author{D.~P.~Marrone}
\affiliation{\Arizona}

\author{J.~J.~McMahon}
\affiliation{\Michigan}

\author{S.~S.~Meyer}
\affiliation{\KICPChicago}
\affiliation{\AAUChicago}
\affiliation{\EFIChicago}
\affiliation{\PhysicsUChicago}

\author{M.~Millea}
\affiliation{\Davis}

\author{L.~M.~Mocanu}
\affiliation{\KICPChicago}
\affiliation{\AAUChicago}

\author{J.~J.~Mohr}
\affiliation{\Munich}
\affiliation{\ExcellenceCluster}
\affiliation{\MPE}

\author{T.~Natoli}
\affiliation{\Dunlap}

\author{S.~Padin}
\affiliation{\KICPChicago}
\affiliation{\AAUChicago}

\author{C.~Pryke}
\affiliation{\Minnesota}

\author{C.~L.~Reichardt}
\affiliation{\Melbourne}

\author{J.~E.~Ruhl}
\affiliation{\CaseWestern}

\author{J.~T.~Sayre}
\affiliation{\CaseWestern}
\affiliation{\Colorado}

\author{K.~K.~Schaffer}
\affiliation{\KICPChicago}
\affiliation{\EFIChicago}
\affiliation{\ArtInstChicago}

\author{E.~Shirokoff}
\affiliation{\Berkeley}
\affiliation{\KICPChicago}
\affiliation{\AAUChicago}

\author{G. Simard}
\affiliation{\McGill}

\author{Z.~Staniszewski}
\affiliation{\CaseWestern}
\affiliation{\JPL}

\author{A.~A.~Stark}
\affiliation{\CfA}

\author{K.~T.~Story}
\affiliation{\KICPChicago}
\affiliation{\PhysicsUChicago}
\affiliation{\KIPAC}
\affiliation{\Stanford}

\author{K.~Vanderlinde}
\affiliation{\Dunlap}
\affiliation{\UToronto}

\author{J.~D.~Vieira}
\affiliation{\Illast}
\affiliation{\Illphys}

\author{R.~Williamson}
\affiliation{\KICPChicago}
\affiliation{\AAUChicago}

\author{W.~L.~K.~Wu}
\affiliation{\KICPChicago}

\collaboration{(The South Pole Telescope Collaboration)}

\correspondingauthor{Ryan Chown}
\email{ryan.chown@mail.mcgill.ca}

\begin{abstract}
We present three maps of the millimeter-wave sky
created by combining data from the South Pole Telescope (SPT) and
the \planck\ satellite.
We use data from the SPT-SZ survey, a survey of 2540 deg$^2$ of the
the sky with arcminute resolution in three bands
centered at 95, 150, and 220 GHz,
and the full-mission \planck\ temperature data in the 100, 143, and 217 GHz bands.
A linear combination of the SPT-SZ and \planck\ data
is computed in spherical harmonic space, with
weights derived from the noise of both instruments.
This weighting scheme results in \planck\ data providing most
of the large-angular-scale information
in the combined maps, with the smaller-scale information coming from SPT-SZ data.
A number of tests have been done on the maps. We find their angular power spectra
to agree very well with theoretically predicted spectra and previously
published results.
\end{abstract}


\section{Introduction} \label{sec:intro}

The most sensitive, highest-resolution all-sky millimeter-wave (mm-wave)
survey was performed by
the \planck\footnote{\url{http://www.cosmos.esa.int/web/planck}}~satellite \citep{planck15-1}.
From 2007 to 2011 the South Pole Telescope\footnote{\url{http://pole.uchicago.edu}} \citep[SPT;][]{carlstrom11} was
used to survey a fraction
of the Southern mm-wave sky
(2540 deg$^2$)
to lower noise levels than the \planck\ full-sky data and with higher resolution ($\sim$1 arcmin).
This survey is referred to as the ``SPT-SZ" survey.
The aim of this paper is to present high-resolution, high signal-to-noise maps
of the mm-wave sky by combining SPT and
\planck\ data in a nearly optimal way.
A map of SPT-SZ data combined with \planck\ will probe, within the SPT-SZ survey area,
mm-wave emission on
both large and small angular scales
with higher signal-to-noise per mode than either SPT-SZ or \planck\ individually.

Away from the Galactic plane and on angular
scales larger than a few arcminutes, the mm-wave
sky is dominated by the cosmic microwave background (CMB).
At small angular scales, individual galaxies are the brightest features:
thermal dust emission from star-forming galaxies \citep[which
make up the cosmic infrared background, or CIB;][]{puget96, gispert00, lagache05}; and synchrotron emission from
active galactic nuclei \citep{dezotti10}.
Inverse Compton scattering of CMB photons by hot intracluster gas
leads to the Sunyaev-Zel'dovich effects, detectable as arcminute-scale
temperature fluctuations or spectral distortions
in the CMB at the positions of galaxy clusters \citep{sunyaev72,sunyaev80}.
The Milky Way is bright at millimeter wavelengths
due to thermal dust emission, synchrotron radiation, and free-free radiation.

\epsscale{0.85}

\begin{figure*}
\plotone{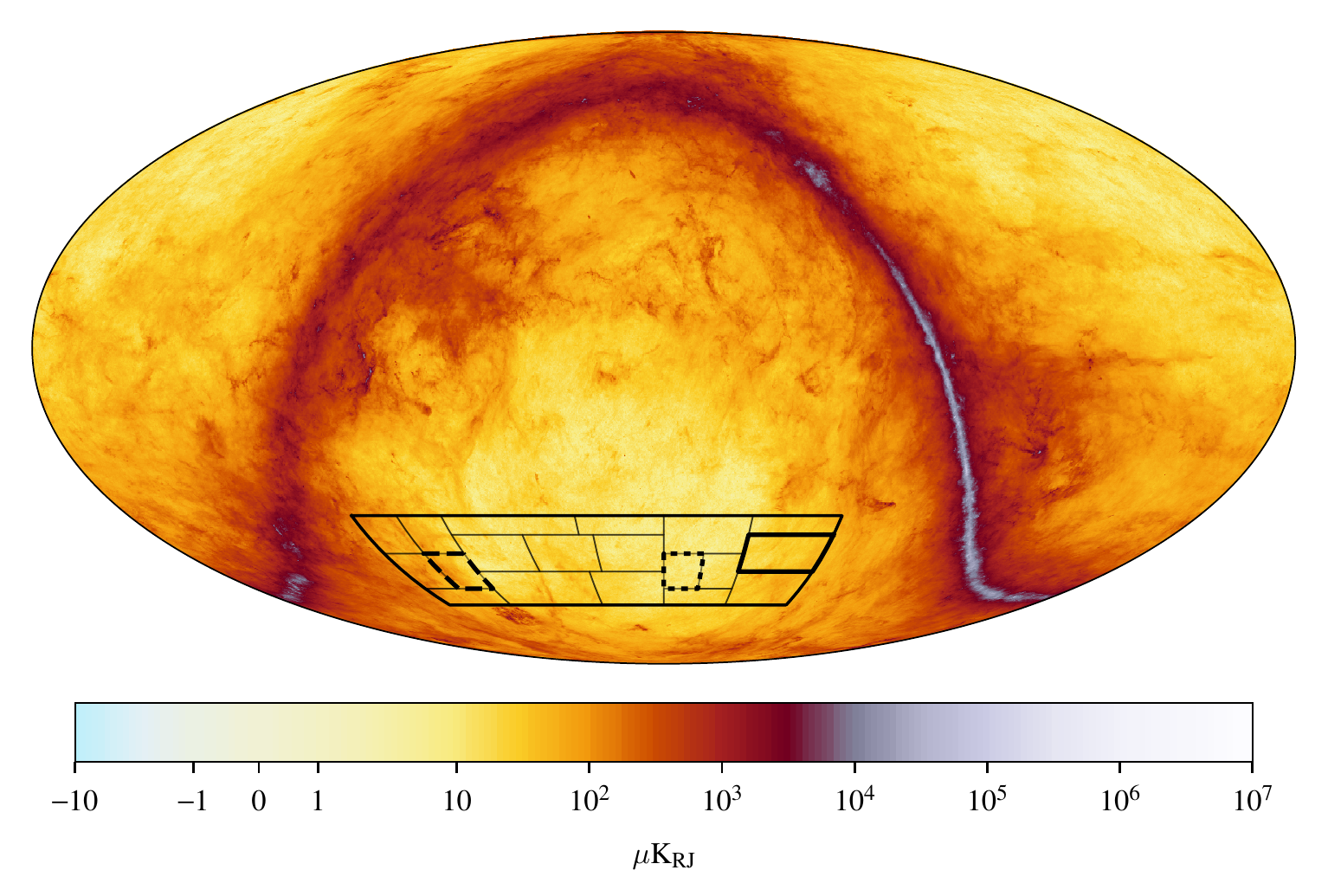}
\caption{
SPT-SZ 2500 deg$^2$ footprint and individual field boundaries
superimposed on the temperature map
of thermal dust from \planck~full-mission data \citep{planck15-10}, shown
in equatorial coordinates in the Mollweide projection. The two fields 
with $\simeq\sqrt{2}$ lower than average noise are
$(\mrm{R.A.},\mrm{dec.})=(5^\mrm{h}30, -55^\circ)$ (dashed boundary), and $(\mrm{R.A.},\mrm{dec.})=(23^\mrm{h}30, -55^\circ)$ (dotted boundary).
The $(\mrm{R.A.},\mrm{dec.})=(21^\mrm{h}, -50^\circ)$ field
(thick solid boundary) has $\simeq\sqrt{2}$ above average noise for this analysis due to an observation cut (see text for details).
The dust map units are
Rayleigh-Jeans brightness temperature at a reference frequency of 545.0 GHz.
\label{fig:2500d_footprint}}
\end{figure*}


Maps of the Large and Small Magellanic Clouds using
a similar combination of
SPT and \planck~temperature maps were presented in \citet{crawford16}.
Similarly to \citet{crawford16}, we use \planck~data to fill in the
information at large angular scales that is
missing from SPT-SZ data and to improve the signal-to-noise at intermediate
angular scales where both instruments have high signal-to-noise.
Meanwhile the higher-resolution SPT data probes small scales
where \planck~is dominated by noise.
We present three maps of the mm-wave sky combined in this way, namely
SPT 95 GHz (3.2 mm) + \planck~100 GHz (3.0 mm),
SPT 150 GHz (2.0 mm) + \planck~143 GHz (2.1 mm), and
SPT 220 GHz (1.4 mm) + \planck~217 GHz (1.4 mm).
Each of these maps cover roughly 2500 deg$^2$ of the Southern sky.
The wide range of angular scales with high signal-to-noise in these maps
makes them useful for a wide array of applications, including
CMB lensing measurements \citep{omori2017, simard2017}.

The structure of this paper is as follows. In Section~\ref{sec:sptintro} and
Section~\ref{sec:planckintro} we introduce the SPT and \planck~instruments and data.
In Sections~\ref{sec:beamfilt} and~\ref{sec:combining} we
describe the filtering, data processing, and the combining procedure.
In Section~\ref{sec:results} we show the resulting combined maps and
present tests of them.
We conclude in Section~\ref{sec:conclusions}.

\section{The South Pole Telescope}\label{sec:sptintro}

The SPT is a 10~m diameter telescope located
at the Amundsen-Scott South Pole Station, Antarctica.
It was constructed primarily to measure
fluctuations in the CMB with high resolution,
and for the detection of galaxy clusters through
their SZ signatures \citep{carlstrom11}.
From 2007 to 2011 a
region of the southern sky spanning 20$^\mrm{h}$ to 7$^\mrm{h}$ in
right ascension (R.A.)
and $-65^\circ$ to $-40^\circ$ in declination (dec.) was observed
in three bands centered at 95, 150, and 220 GHz,
with resolutions of approximately 1.7, 1.2, and 1.0 arcmin, respectively.
This patch of the sky contains relatively low levels of Galactic
foreground emission, as shown in comparison to the
thermal dust map from \planck~data \citep{planck15-10} in Figure~\ref{fig:2500d_footprint}.
The observations were performed in nineteen sub-regions which together comprise a
contiguous 2540 deg$^2$ area on the sky \citep[][hereafter S13]{story13}
referred to as the ``SPT-SZ survey."


In this paper we use 95 GHz observations taken between
2009-2011, 150 GHz observations from 2008-2011, and
220 GHz observations from 2008-2011.
Two of the fields 
were observed for roughly twice the average amount of time in the 150 and 220 GHz bands and hence have lower than average noise. These fields are 
indicated in Figure~\ref{fig:2500d_footprint}.
Roughly half of the $(\mrm{R.A.},\mrm{dec.})=(21^\mrm{h}, -50^\circ)$ observations
(thick solid outline in Fig.~\ref{fig:2500d_footprint}) employed a
different scanning strategy and are not used here,
making this field noisier than average for this analysis.
We use the same 2008-2011 observation cuts as Mocanu et al. 2018 (in preparation),
except for the 150 GHz $(23^\mrm{h}30, -55^\circ)$ field, where we use the observation
cuts from S13.



\section{The \textit{Planck} Satellite}\label{sec:planckintro}

The \planck~satellite completed approximately four and a half surveys of
the entire sky (one every six months) between August 2009 and October 2013 \citep{planck15-1}.
The \planck~High-Frequency Instrument (HFI) observed the sky in six frequency bands
centered at 100, 143, 217, 353, 545, and 857 GHz.
Maps of the sky in these bands
were released in 2013 \citep{planck13-1} and in the 2015 Public Release 2
\citep[PR2;][]{planck15-1} with greater sensitivity and improved calibration accuracy.
We use the PR2 full-mission HFI maps from the 100, 143, and 217 GHz bands, which are those
closest in frequency to the SPT-SZ bands (Figure~\ref{fig:beams}b).
The resolution of the 100, 143, and 217 GHz \planck~maps are
approximately 10.0, 7.1, and 5.0 arcmin, respectively.

\section{Response Functions and Data Processing}\label{sec:beamfilt}

In order to combine SPT data with \planck~data we must
deconvolve their response functions due to
instrument beams and filtering.
In this section we describe the model for the total beam-and-filtering
response functions.
For SPT we describe the steps of filtering and the calculation of the
filter transfer functions for each SPT-SZ band.
We use the approximation that for each band, a single two-dimensional
transfer function describes the filtering over the
entire 2500 deg$^2$ area.
The Planck Collaboration has calculated and published their
total response functions, so we only briefly describe them here.

A map of temperature fluctuations on the sphere $\Delta T(\vec{\theta})$ multiplied by
a mask $M(\vec{\theta})$
may be decomposed into spherical harmonic coefficients $\tilde{a}_{\ell m}$ using the
spherical harmonic transform:
\begin{equation}\label{eq:sht}
\tilde{a}_{\ell m} = \int_S \Delta T(\vec{\theta}) M(\vec{\theta}) Y_{\ell m}(\vec{\theta}) \dee \Omega,
\end{equation}
where the tilde on $\tilde{a}_{\ell m}$ indicates that mode coupling due to the
application of the mask has not been corrected for.
%

The action of a beam response function $B_{\ell m}$ 
and a filter transfer function $t_{\ell m}$ on the 
spherical harmonic coefficients of the true temperature map $\tilde{a}_{\ell m}^\mrm{in}$ 
yields the spherical harmonic coefficients of the observed temperature map 
$\tilde{a}_{\ell m}^\mrm{out}$.
This can be written as
\begin{equation}
\tilde{a}_{\ell m}^\mrm{out} = t_{\ell m} B_{\ell m} \tilde{a}_{\ell m}^\mrm{in}.
\end{equation}
Note that $\tilde{a}_{\ell m}^\mrm{in}$ (with the tilde) are the
harmonic coefficients of the true full-sky temperature map $a_{\ell m}^\mrm{in}$
(no tilde) after applying the survey mask.
From here, the total response function $F_{\ell m}$ is defined:
\begin{align*}
F_{\ell m} & \equiv t_{\ell m} B_{\ell m} \\
& = \frac{\tilde{a}_{\ell m}^\mrm{out}}{\tilde{a}_{\ell m}^\mrm{in}}.
\end{align*}


\subsection{SPT}

\begin{figure*}
\begin{centering}
\gridline{\fig{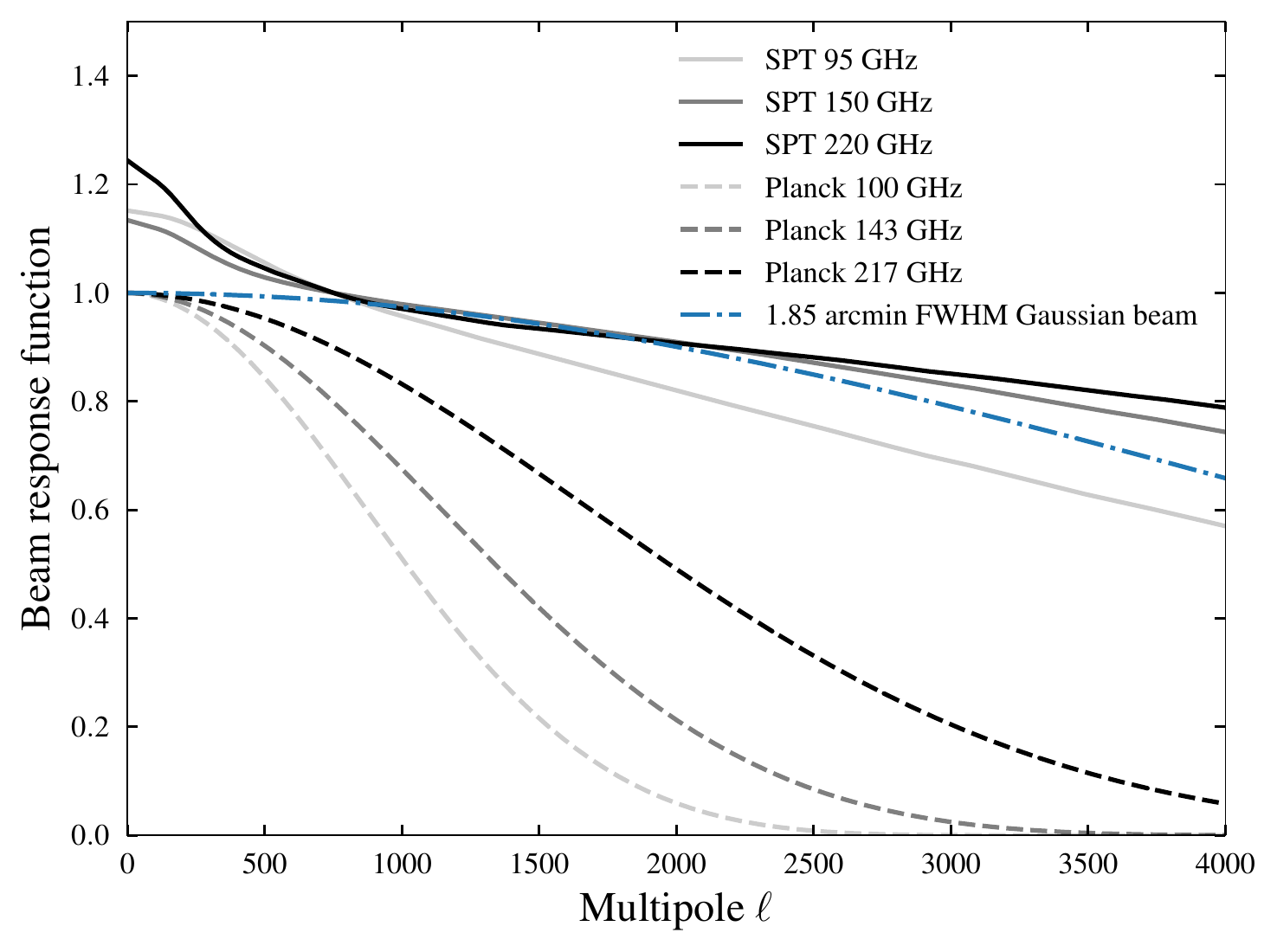}{0.5\textwidth}{(a) }
		  \fig{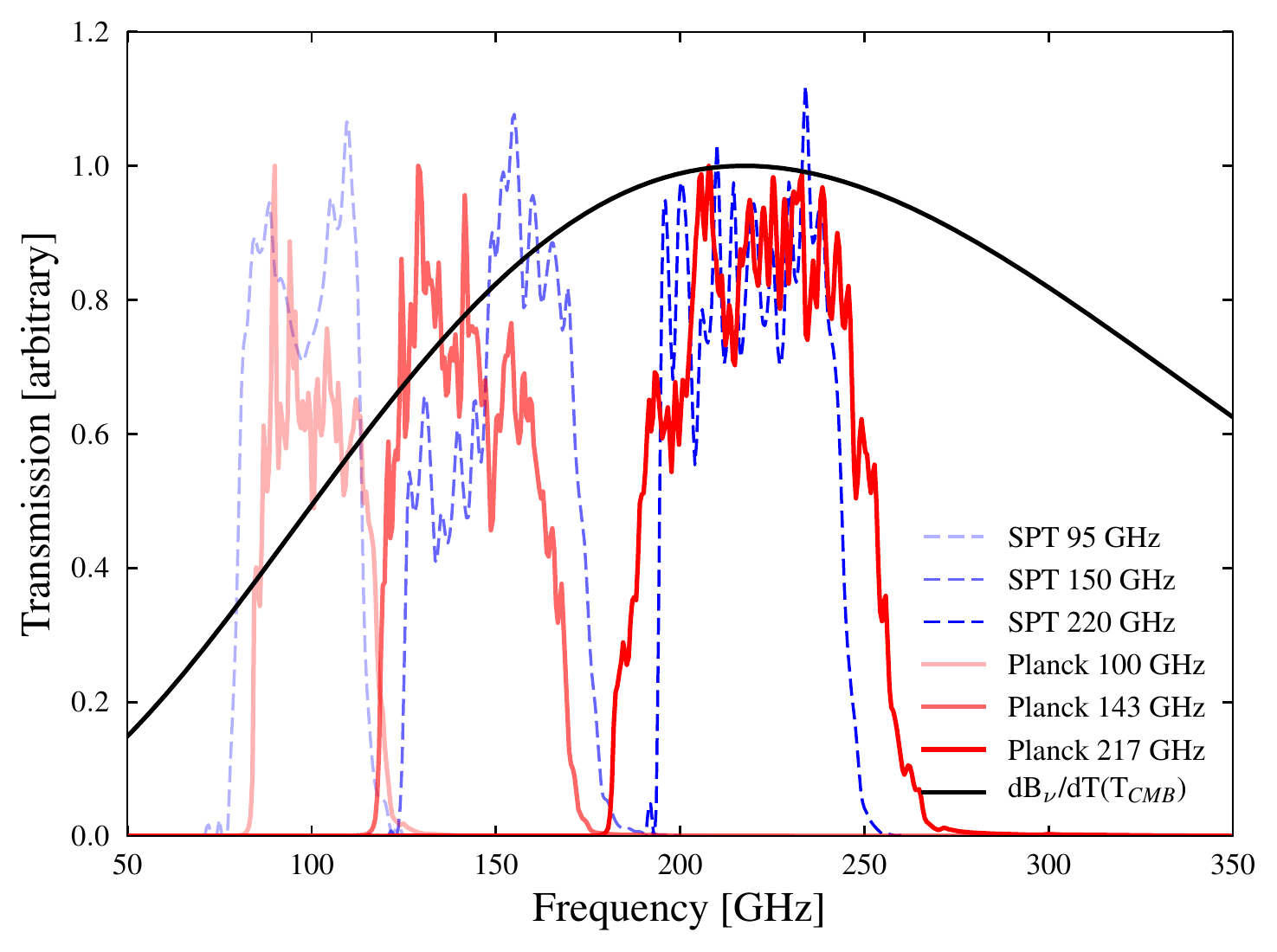}{0.5\textwidth}{(b) }
}
\caption{(a) Year-averaged SPT beams for each band \citep[using individual year beams from][]{keisler11}. The SPT beams exceed 1.0 because they are normalized to 1.0 at $\ell = 750$.
Also shown are the \planck~beam window functions (which include the beam \textit{and} filtering) from the \planck~Reduced Instrument Model, version 2.00 \citep{planck15-7} as well as the final beam ($B_\ell^\mrm{final}$, in Section~\ref{sec:combining}) we apply to the combined data maps (in blue).
(b) Transmission bandpass functions for SPT and the nearest \planck~HFI bands, and the
derivative of the blackbody spectrum d$B_\nu$/d$T$ evaluated at $T_\mrm{CMB}$ overlaid in black.
\label{fig:beams}}
\end{centering}
\end{figure*}

The beam response of the SPT is well approximated as a circularly symmetric beam $B_\ell$, and
was estimated with percent-level precision using measurements of planets and
bright extragalactic sources, as detailed in, e.g., \citet{schaffer11,keisler11}.
Over the four years of data presented here, the SPT optics were modified slightly, and
the observing frequency distribution of detectors in the focal plane changed; as a result,
the instrument beams are slightly different for different observing seasons. In this work,
we use beams averaged over the four years, with inverse-noise weighting.
These year-averaged beams for each band
are plotted in Figure~\ref{fig:beams}.


The filter response of SPT in past analyses has been computed
for each individual field and each frequency in flat-sky coordinates.
In this analysis, for each band, we combine all of the fields into a contiguous
2500 deg$^2$ map, and we compute a two-dimensional transfer function
to characterize the filtering of the full 2500 deg$^2$ survey.
In the following sections we describe these filtering steps and
the calculation of the filter transfer functions.

\subsubsection{Time Stream Filtering}\label{sec:timestream}

The SPT was used to observe each field in a series of observations which are composed of
successive scans across the sky along azimuth. Adjacent scans are
separated by a small step in elevation.
As the telescope is located at the South Pole, the scan direction is along R.A.,
with steps in declination.
Time-varying emission from the atmosphere leads to increased noise
on large angular scales.
The raw time stream data in each scan is filtered to reduce
this noise. For more detail on SPT-SZ filtering, see \citet{schaffer11}.

The time stream data from each scan are fit with a
seventh-order polynomial 
and set of low-order sines and cosines, which are subtracted from these data.
This results in an effective scan-direction high-pass filter with a cutoff of $\ell \simeq 270$.
Sources measured to be brighter than 50 mJy at 150 GHz are excluded from the fitting
(with a 5-arcmin masking radius).
Fainter sources are not masked before filtering, which gives them wing-like features (aligned
with the scan direction) in the resulting maps.

The six modules of the SPT-SZ camera each contain 160 detectors. Each module 
is equipped with filters determining their observing 
frequency (95 GHz, 150 GHz, or 220 GHz).
At every time sample and separately for each module, 
the mean and two spatial gradients of the data 
from all detectors in a module is subtracted from their data. 
This reduces atmospheric noise at large angular scales.
A low-pass filter is applied in Fourier space to the data from each
detector to avoid aliasing when the data are sampled into map pixels.

Note that the transfer function $t_{\ell m}$ as defined in
Equation~\ref{eq:tlm} is an approximation when applied to SPT-SZ data, in that
the actual filtering is performed on individual scans in the time domain
and in Fourier space, which does not transform perfectly into a simple convolution.
However, as we will show in Section~\ref{sec:powerspectra},
this is a very good approximation.

\subsubsection{Constructing \hpx~Maps from SPT Fields}\label{sec:2500d_interp}

We use the ``Hierarchical Equal Area isoLatitude Pixelation of a sphere"
\citep[\hpx;][]{gorski05}\footnote{\url{http://healpix.sourceforge.net}}
scheme
to pixelize our maps.
The filtered SPT maps of each field are initially in the
Lambert azimuthal equal-area projection. 
We match the response of each field from the year-varying beams
into a common beam in two-dimensional Fourier space. The common beam
is chosen to be a $\simeq 2$ arcmin full width at half maximum (FWHM) Gaussian,
however this is not the final resolution of the combined maps. This common beam
is replaced at the end of the combining pipeline with a 1.85 arcmin FWHM Gaussian.

The fields are beam matched by multiplying the
two-dimensional FFT of the co-added temperature map of each field with the ratio of
the common 
beam to the beam for that year and frequency.
Then in position space we perform bilinear interpolation of these beam-matched
fields onto the nearest \hpx~pixel locations (with resolution $\nside=8192$).
We interpolate the weights for each field onto the same \hpx~grid, and compute
the weighted sum of the temperature values using these weights.
The final combined maps are in \hpx~format with resolution $\nside=8192$.

\subsubsection{Masking}\label{sec:masking}

We mask the bright sources (flux densities greater than 50.0 mJy at 150 GHz) which are masked during time stream filtering,
and regions close to the SPT-SZ boundary.
This mask is constructed by cutting holes with radius of 5 arcmin in an SPT-SZ
boundary mask. We apodize the mask outside the holes and at the boundary
with a Gaussian with $\sigma=5$ arcmin.
This mask is applied to SPT and \planck~data, noise and simulations.

\subsubsection{Calculating the Transfer Functions}\label{sec:tfcalc}

The filter transfer functions for each of the 2500 deg$^2$ SPT-SZ maps are computed
as follows.
We create 100 full-sky simulated input maps per band consisting of
the lensed CMB,
Gaussian foregrounds (which are correlated between bands),
and the Poisson-distributed population of point sources with 150 GHz flux densities
in the range $6.4~\mathrm{mJy} \leq F_{150} \leq 50.0~\mathrm{mJy}$ (Everett et al., in preparation).
Simulated lensed CMB maps are generated by running LensPix \citep{lewis05} on
temperature power spectra
derived from the \textsc{\planck\ TT + lowP + lensing} cosmology\footnote{{\tt base\_plikHM\_TT\_lowTEB\_lensing}} \citep{planck15-13}.
We simulate SZ-detected galaxy clusters
with statistical significance $\xi\geq 5.0$ in the 150 GHz band from
the \citet{bleem2015} catalog.
The clusters are
simulated in their observed positions in the 95 GHz and 150 GHz bands (there
is negligible SZ signal at 220 GHz).
Using the integrated
Comptonization $Y_\mrm{SZ}$ (over an 0.75 arcmin radius disk)
and core radius $\theta_c$ taken
from the \citet{bleem2015} catalogue, the radial temperature profile of
each cluster $\Delta T(\theta)$
is computed assuming a projected isothermal
$\beta$-model \citep{cavaliere76} with $\beta=1$:
\beq
\Delta T(\theta) = y_0 T_\mrm{CMB} g_\nu (1+\theta^2/\theta_c^2)^{\frac{1-3\beta}{2}},
\eeq
where $y_0$ is the peak Compton $y$-parameter, the CMB temperature $\tcmb=2.7255$ K
\citep{fixsen09b}, $g_\nu$ is a function of observing frequency $\nu$,
$\theta$ is the angular distance from the
cluster center, and $\theta_c$ is the cluster core radius \citep{bleem2015}.

We create 100 simulated maps (one per input sky) for every individual observation of each field at each band. We create mock timestreams from each input map, and these mock timestreams are made into single-observation simulated maps using the same filtering and processing as used on the real data. The individual observation maps of each field at each band are then coadded, and the single-field maps are combined into full 2500 deg$^2$ maps in the same way as the real data. This results in 100 simulated 2500 deg$^2$ maps in each band; these maps have the same statistical properties as our best estimate of the observed sky.

\begin{figure*}
\gridline{\fig{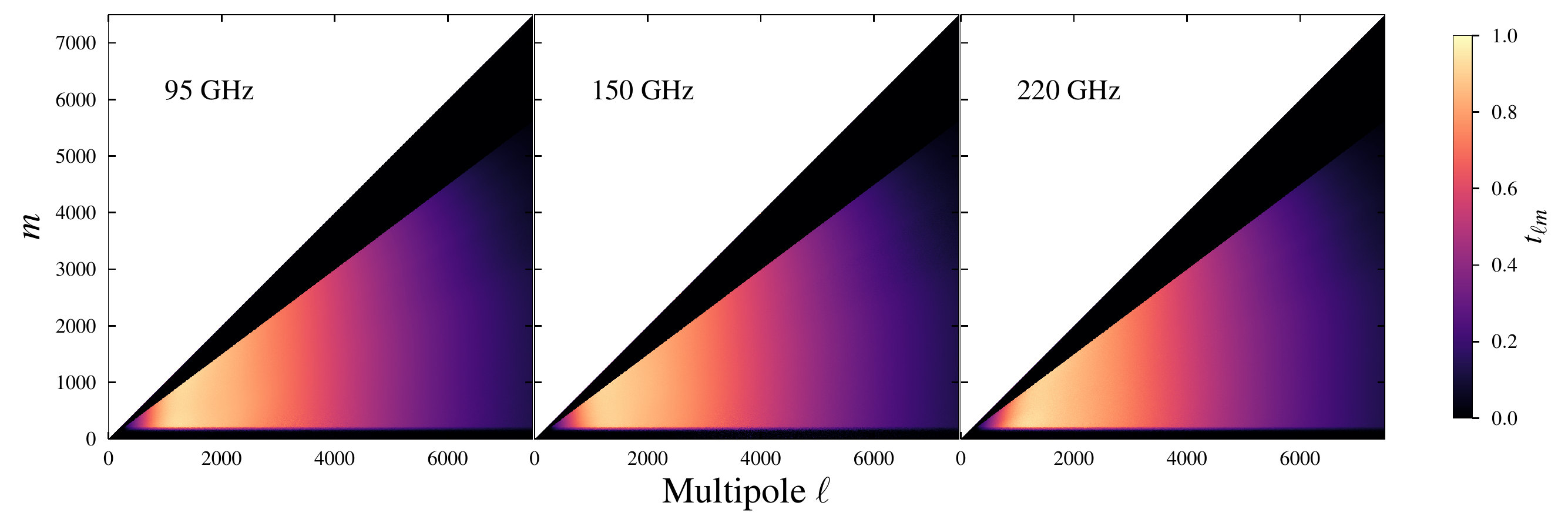}{\textwidth}{}}
\caption{ Two-dimensional filter transfer functions $t_{\ell m}$ of 2500 deg$^2$ SPT-SZ data.
They are plotted up to $\ell=8000$ to make the $m\lesssim 300$ filtering visible,
however in the analysis we have calculated them up to $\ell = 16000$.
Smaller values of $t_{\ell m}$ indicate more strongly-filtered modes.
The common beam from beam-matching the fields is not included in these plots; the
suppression of power with increasing $\ell$ is due to the 1 arcmin pixelization.
The dark wedge of modes at high $m$ have been set to zero, since these modes have been 
strongly suppressed due to the mask (see explanation in text) and contribute negligibly to the combined maps.
Note that the \planck~beam and filter response functions, shown in
Figure~\ref{fig:beams}a, are independent of $m$, so there is no need to show them here as well.
\label{fig:2500d_tfs}}
\end{figure*}

The input maps and observed maps
(in \hpx~format using the method described in Section~\ref{sec:2500d_interp}) are
masked, and their spherical harmonic coefficients are
computed ($\tilde{a}_{\ell m}^{\mrm{in}}$ and $\tilde{a}_{\ell m}^{\mrm{out}}$, respectively).
The filter transfer function is then computed by
\begin{equation} \label{eq:tlm}
t_{\ell m} \equiv \frac{1}{B_\ell} \frac{\left\langle \tilde{a}_{\ell m}^{\mrm{out}}\left(\tilde{a}_{\ell m}^{\mrm{in}}\right)^*\right\rangle}{\left\langle \tilde{a}_{\ell m}^{\mrm{in}}\left(\tilde{a}_{\ell m}^{\mrm{in}}\right)^*\right\rangle},
\end{equation}
where $\langle ... \rangle$ denotes the ensemble average over all simulations, and
$B_\ell$ is the common beam 
introduced in Section~\ref{sec:2500d_interp}.
Averaging over 100 simulations per band,
we compute the transfer function using this equation.
The filter transfer functions for each SPT-SZ band are plotted
in Figure~\ref{fig:2500d_tfs}.
Applying the property of spherical harmonic coefficients
$a_{\ell (-m)} = a_{\ell (+m)}^*$ to $t_{\ell m}$, whose imaginary part is negligible,
implies $t_{\ell (-m)} = t_{\ell (+m)}$. We therefore only need to plot
half of the total number of modes (e.g. $m\geq 0$ or $m\leq 0$).
One can interpret $t_{\ell m}$ with $\ell$ on the x-axis and
$m$ on the y-axis as follows:
\begin{itemize}
\item Larger values of $m$ correspond to smaller angular scales
along the SPT scan direction in equatorial coordinates.
\item Filtering in the time domain leads to the localized strip of
suppressed modes in $t_{\ell m}$ at $m \lesssim 300$.
\item Applying a mask
with a restricted range of declination and positioned
away from the equator (such as the SPT-SZ
survey mask) to a full-sky map
leads to a triangle-shaped region of suppressed modes at high $m$.
This feature is not due to filtering, it is purely from
the geometry of the survey mask when viewed in spherical harmonic space. 
High-$m$ spherical harmonics have most of their power at the equator, so the survey mask suppresses these modes.
\item The $\ell$-dependent trend in $t_{\ell m}$ is the smoothing due to the
1 arcmin pixelization, and is the same for all three frequencies.
Note that we are showing $t_{\ell m}$ exactly as in Equation~\ref{eq:tlm}, so the
$\simeq 2$ arcmin beam from beam-matching $B_\ell$ is not included.
\end{itemize}

\subsubsection{Calibration}

The units of CMB maps are differential temperature relative to the CMB temperature,
\beq
\Delta T \equiv T - \tcmb,
\eeq
denoted K$_\mrm{CMB}$.
The calibration accuracy of the \planck~maps in the 100, 143, and 217 GHz bands
are $\simeq 0.1$ per cent \citep[Table 6 of][]{planck15-8}.
We use this highly precise \planck\ calibration to calibrate the SPT-SZ maps. We use the 150 GHz
calibration factor derived from comparison to \planck\ 143 GHz data in the SPT-SZ footprint from \citet{hou2018}. We then inter-compare the SPT-SZ 95, 150, and 220 GHz maps to obtain calibration factors at 95 and 220 GHz.
The uncertainty of these relative calibrations are 0.22, 0.15, and 0.43 percent (in map units) at
95, 150 and 220 GHz respectively.

\subsubsection{Beam and Filter Deconvolution}\label{sec:tfdeconv}

Upon calculating the spherical harmonic coefficients of the SPT data
$\tilde{a}_{\ell m}^{\mrm{SPT}}$ we divide out the filter transfer function $t_{\ell m}$
and the common beam $B_\ell$.
For the small subset of modes at low $m$ which are heavily suppressed due to
filtering in the time domain, the transfer function is close to zero and
there is negligible signal.
We do not deconvolve $t_{\ell m}$ for modes where $m<300$.
These low-$m$ modes get filled in with \planck~data later on.
The transfer function also becomes exponentially small at high $m$ due to
a low-pass filter. This filter is not the reason for the triangle-shaped wedge
of modes at high $m$ --  that is due to the application of the 2500 deg$^2$ survey mask -- but
it causes the transfer function to fall off at $\ell\simeq8000$, $m\simeq5000$.
Modes with $m>5000$ remain filtered in the final maps.

\subsubsection{Noise Estimation} \label{sec:noise_estimation}

A half-difference map is calculated for each observation of each field, such
that the left-going scans (along increasing R.A.) and right-going scans
(along decreasing R.A.) are differenced and divided by 2. This nulls the signal in each
scan while preserving the statistics of the noise.
The mean of all of the half-difference maps for a field, with a randomly selected half of
them multiplied by $-1$, gives us an estimate of the noise in that field.
Using random $\pm 1$'s allows us to produce more realizations of the noise.
We make 100 noise realizations of each field and each frequency, and construct
a \hpx~map out of each of them.

\begin{figure*}
\gridline{\fig{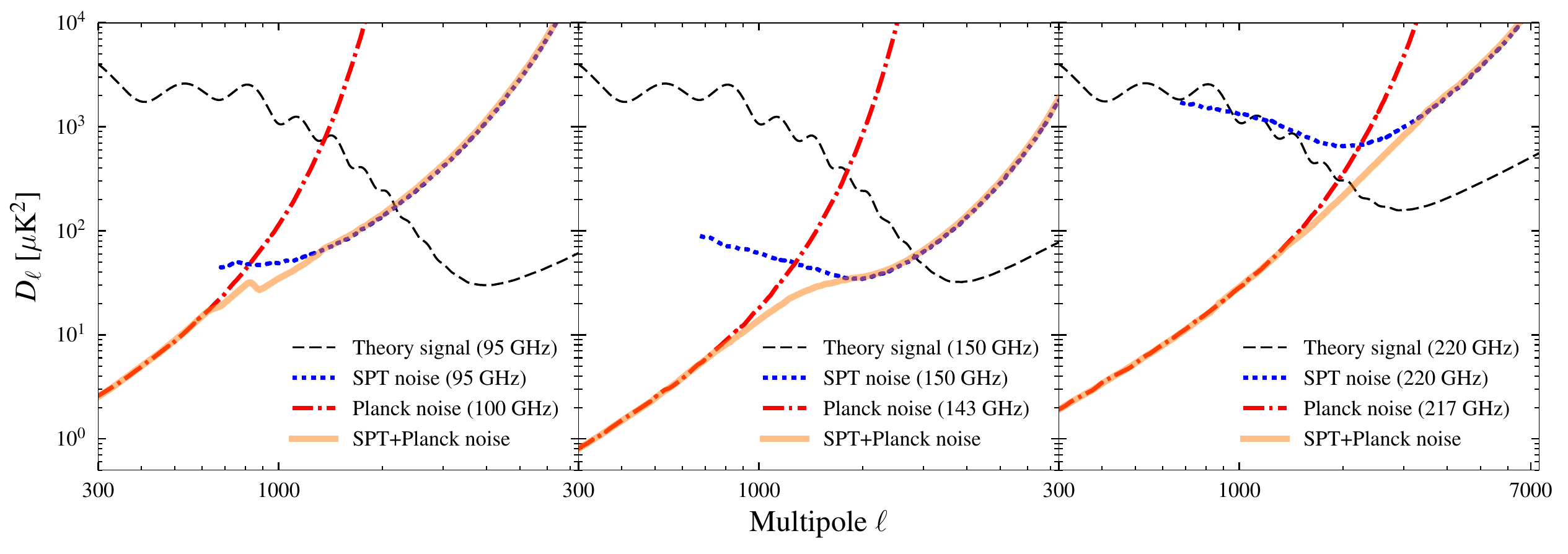}{\textwidth}{}}
\caption{ Average auto-spectra of noise realizations of SPT-SZ (dotted), \planck\ (dot-dashed),
and combined SPT-\planck\ (solid), in comparison with the total expected signal
(CMB plus foregrounds in the 2500 deg$^2$ patch; the dashed lines). Left panel = 95/100 GHz, middle panel = 150/143 GHz, right panel = 220/217 GHz. The $m<300$ modes were down-weighted
with the same filter as the data.
\label{fig:2500d_noise_1d}}
\end{figure*}

To estimate the two-dimensional noise power $N_{\ell m}$ in each band,
we apply the boundary and point-source mask to each noise realization,
compute their harmonic transforms $\tilde{n}_{\ell m}$, deconvolve the beam and filter response functions,
and then compute their variance:
\begin{equation}\label{eq:noisevar}
N_{\ell m} = (F_{\ell m})^{-2}\left\langle\left|\tilde{n}_{\ell m}\right|^2\right\rangle.
\end{equation}
The auto-spectrum of the SPT noise realizations for each frequency, with their beam and filter transfer functions deconvolved, are shown by the dotted lines in Figure~\ref{fig:2500d_noise_1d}.

%
%

\subsection{\planck}

The total beam-and-filter response functions $F_{\ell m}$ for the \planck~instrument --
referred to as ``beam window functions" in the \planck~literature --
have been computed by the Planck Collaboration and are available
online through the Planck Legacy Archive\footnote{\url{https://www.cosmos.esa.int/web/planck/pla}}.
The beam window functions for each \planck~band are
well-approximated to depend only on $\ell$ (i.e. $F_{\ell m}=F_\ell$);
they are plotted in Figure~\ref{fig:beams}.
The beam window functions do not include the smoothing due to $\nside=2048$ pixelization;
we account for this separately.

We use the full-mission \planck~HFI maps, which are $\nside=2048$ resolution
in Galactic coordinates. We compute the spherical harmonic coefficients
of these maps (without any masking) and rotate them to equatorial coordinates
using the \texttt{rotate\_alm} \hpx~function.
We then invert the spherical harmonic transform and mask the resulting maps
using the mask from Section~\ref{sec:masking}.
We compute the spherical harmonic transform of the masked \planck~maps,
divide out the beam window functions, and multiply by the ratio of the $\nside = 2048$
to $\nside=8192$ pixel window functions to match the smoothing due to pixelization
with the SPT maps.

For each \planck~band we use 100 of the ``8th Full Focal Plane"
noise realizations from the \planck\ 2015 data release 
\citep[see][]{planck15-12} obtained from the Planck Legacy Archive.
These are designed to mimic the true noise statistics (including spatial variation)
in the full-mission data maps.
We repeat the same processing steps as the real data on these \planck~noise realizations,
then compute the two-dimensional noise power using Equation~\ref{eq:noisevar}.
The auto-spectrum of the \planck\ noise realizations for each frequency, with their 
beam and filter transfer functions deconvolved, are the dot-dashed lines 
in Figure~\ref{fig:2500d_noise_1d}.

\section{Inverse Noise-power Weighting in 2-D Harmonic Space}\label{sec:combining}

The end result of the previous section are the deconvolved spherical harmonic
coefficients of SPT $\tilde{a}_{\ell m}^\mrm{SPT}$ and \planck~data $\tilde{a}_{\ell m}^\mrm{\planck}$
for each SPT and \planck~observing band;
and estimates of the noise power at each mode for SPT $N_{\ell m}^\mrm{SPT}$ and
\planck~$N_{\ell m}^\mrm{\planck}$, also in each observing band.

\begin{figure*}
\gridline{\fig{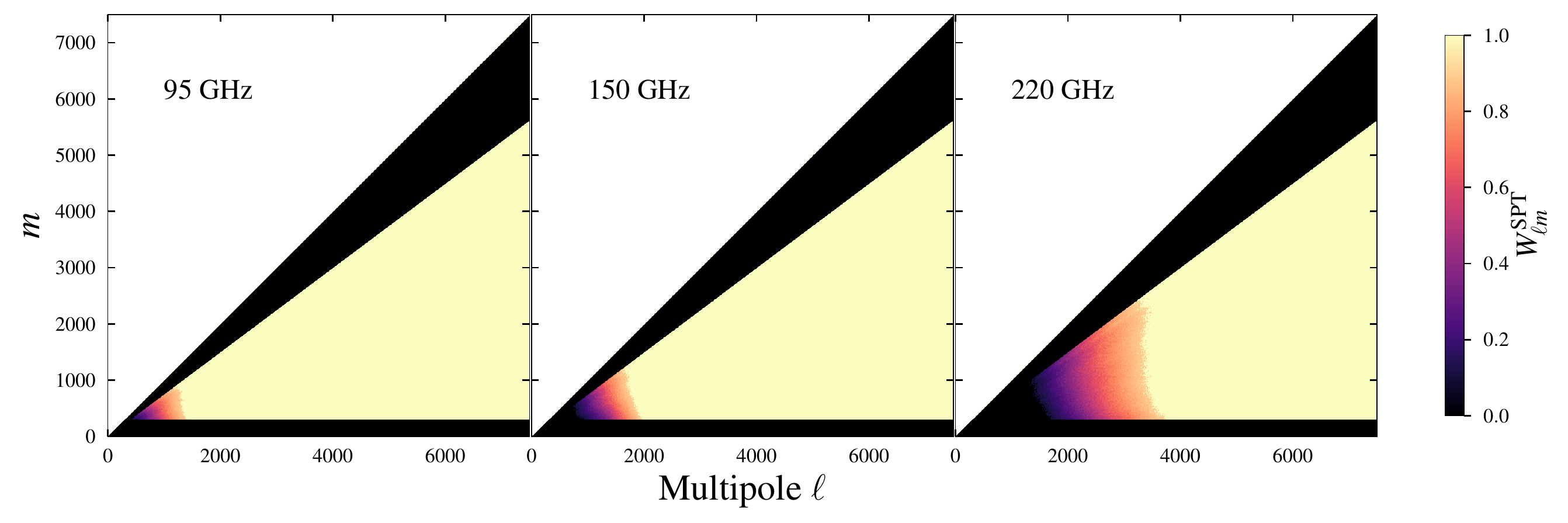}{\textwidth}{(a) SPT weights $W_{\ell m}^\mrm{SPT}$}
}
\gridline{\fig{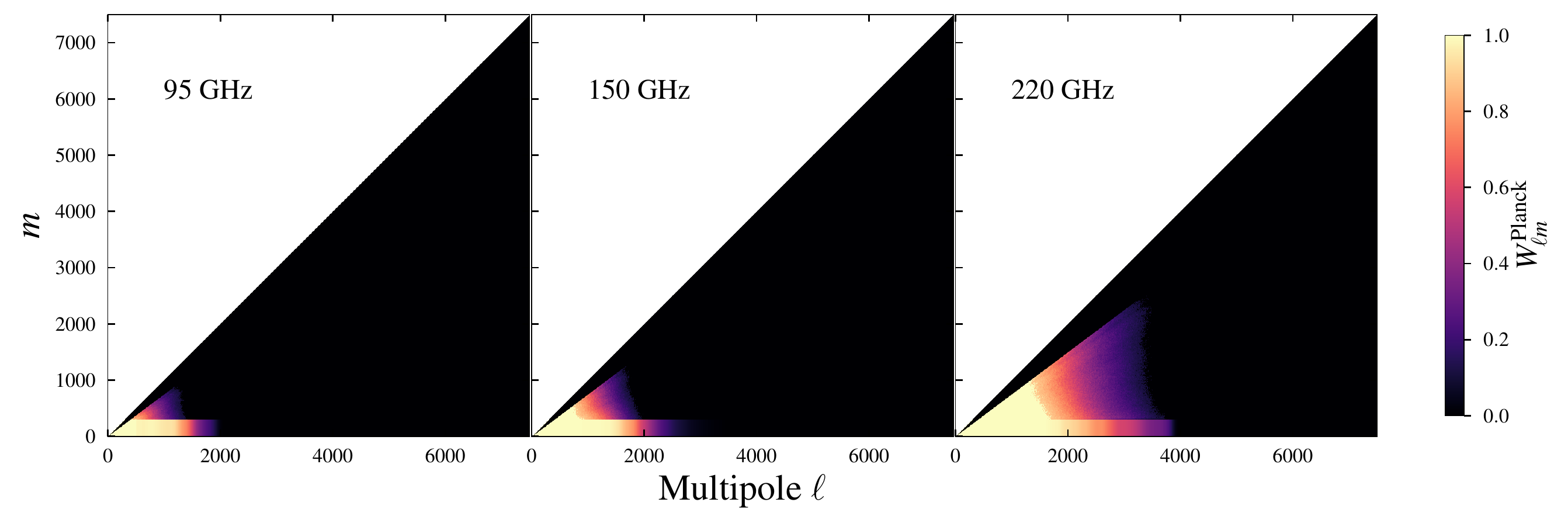}{\textwidth}{(b) \planck~weights $W_{\ell m}^\mrm{Planck}$}
}
\caption{Two-dimensional inverse noise-power weights for (a) SPT  (equation~\ref{eq:wlmspt}) and (b) \planck\ (equation~\ref{eq:wlmplanck}).
All weights have been multiplied by the filter $M_{\ell m}$ for each band,
which is why both SPT and \planck\ weights are very small at high-$\ell$, low-$m$
(i.e. they do not sum to 1.0 where $M_{\ell m} \neq 1.0$).
The dark wedge of modes at high $m$ have been set to zero, since these modes have been 
strongly suppressed due to the mask and contribute negligibly to the combined maps.
\label{fig:2500d_noise}}
\end{figure*}

The data are combined through a linear weighting in two-dimensional
harmonic space
\begin{equation} \label{eq:combining}
\tilde{a}_{\ell m}^\mrm{SPT+\planck} = \tilde{a}_{\ell m}^\mrm{\planck} W_{\ell m}^\mrm{\planck} + \tilde{a}_{\ell m}^\mrm{SPT}W_{\ell m}^\mrm{SPT},
\end{equation}
where $W_{\ell m}^\mrm{SPT}$ and $W_{\ell m}^\mrm{\planck}$ are the (real-valued)
weights.
We choose to construct weights which
minimize the noise power at each mode in the resulting maps.
These weights are given by
\begin{equation}\label{eq:wlmspt}
W_{\ell m}^\mrm{SPT} \equiv \frac{\left(N_{\ell m}^{\mrm{SPT}}\right)^{-1}}{\left(N_{\ell m}^{\mrm{SPT}}\right)^{-1}+\left(N_{\ell m}^{\mrm{\planck}}\right)^{-1}},
\end{equation} and
\begin{equation}\label{eq:wlmplanck}
W_{\ell m}^\mrm{\planck} \equiv \frac{\left(N_{\ell m}^{\mrm{\planck}}\right)^{-1}}{\left(N_{\ell m}^{\mrm{SPT}}\right)^{-1}+\left(N_{\ell m}^{\mrm{\planck}}\right)^{-1}}.
\end{equation}
These weights are shown in Figure~\ref{fig:2500d_noise}. 
One can see that the $\ell$-dependence of the weights (Figure~\ref{fig:2500d_noise})
qualitatively agrees with the $\ell$-dependence of the 
SPT-only and \planck-only noise power spectra in Figure~\ref{fig:2500d_noise_1d}; 
the \planck\ weights are close to 1.0 at low $\ell$ (where \planck\ noise is lower than SPT), 
and close to 0.0 at high $\ell$ (where \planck\ noise is much greater than SPT), 
and vice-versa for the SPT weights.

As previously mentioned, 
the subset of modes at $m\lesssim 300$ are exceptional in that the SPT power has been
completely removed by filtering,
while the \planck\ noise power (and hence the combined map noise power at low $m$) increases
steeply with $\ell$ (roughly $\sim B_\ell^{-2}$) after deconvolving the \planck\ beam window
function $B_\ell$. 
This leads to the total power (signal+noise) per mode $|\tilde{a}_{\ell m}|^2$
being up to orders of magnitude greater at low $m$ than at high $m$ for a given $\ell$.
In other words, the noise in the combined maps at low $m$ (contributed by \planck) 
becomes the dominant contribution to the total map power at high $\ell$.
This is a problem for the visual appearance of the combined maps, and it negatively
affects the stability of the algorithm we use to calculate the inverse spherical
harmonic transform (\texttt{alm2map}). We choose to filter this small subset of modes
in the combined maps such that at fixed $\ell$, the average power of the low-$m$
modes is approximately equal to the average power of the higher-$m$ modes.
We do this by multiplying $\tilde{a}_{\ell m}^\mrm{SPT+\planck}$ of the combined data by a filter $M_{\ell m}$
defined to be 1.0 for all $\ell$ and $m$ except the region encompassing $m<300$ for $\ell>600$, where it is
set to
\beq
M_{\ell m} \equiv \left\langle\frac{\sum_{m=350}^{650}|\tilde{a}_{\ell m}^\mrm{SPT+\planck(sim)}|^2}{\sum_{m=0}^{300}|\tilde{a}_{\ell m}^\mrm{SPT+\planck(sim)}|^2}\right\rangle_{\mrm{sims}},
\eeq
where $\tilde{a}_{\ell m}^\mrm{SPT+\planck(sim)}$ are
from noisy simulations.

\begin{figure*}
\gridline{\fig{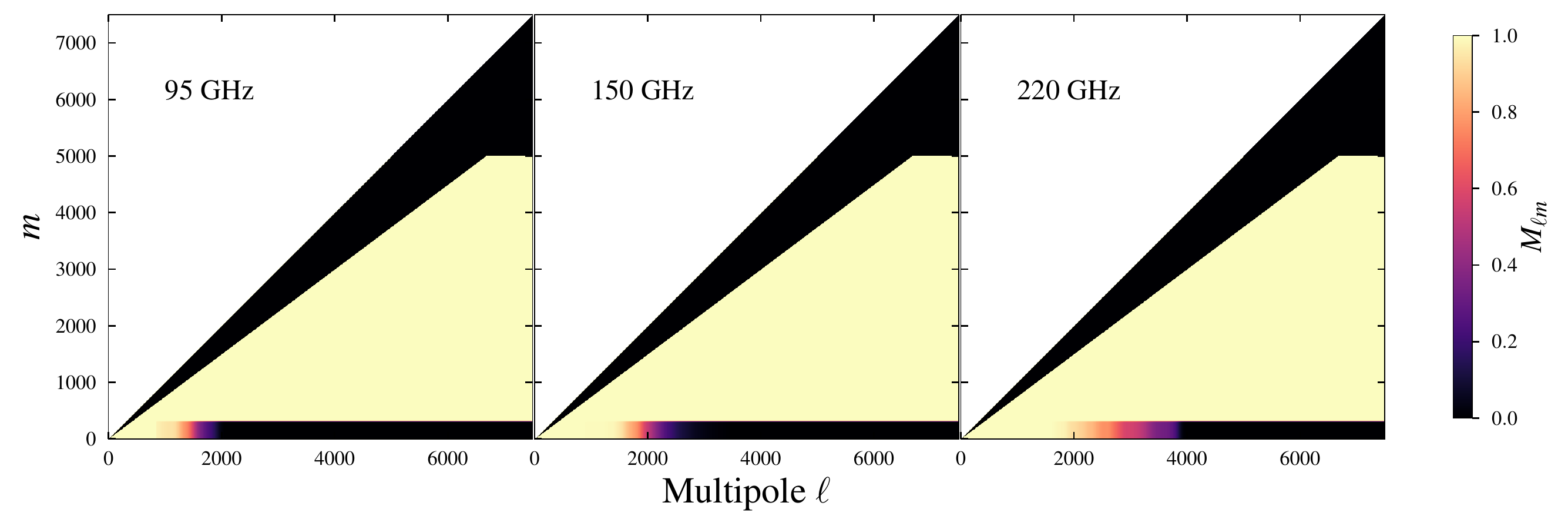}{\textwidth}{}
}
\caption{ The two-dimensional filters $M_{\ell m}$ designed to
ensure uniform power (signal+noise) across $m$ at each $\ell$ in the
combined data maps. We have zeroed out $m\geq 5000$ because SPT data has been 
very heavily suppressed (Section~\ref{sec:tfdeconv}). To maintain consistency with 
previous plots we have zeroed out modes which are heavily suppressed due to the mask 
(the black region at $m \simeq \ell$.).
\label{fig:masklm}}
\end{figure*}

This filter ensures spatially uniform ($m$-independent) signal+noise in the maps.
Additionally, we set $M_{\ell m}$ to zero for $m>5000$ modes; 
above $m \simeq5000$, the SPT filtering has essentially zeroed out any signal,
and we choose not to include these modes in the final maps (as mentioned
in Section~\ref{sec:tfdeconv}).
The filters for each band are shown in Figure~\ref{fig:masklm}.
Note that the application of this filter means that the combined maps are not
truly unbiased (i.e. having all filters deconvolved), however
it improves the visual appearance of the maps by suppressing noisy low-$m$ modes.
This filter will be made publicly available, so that users
may deconvolve it from the maps and use a different filter if they wish.

Finally, we convolve with a final beam $B_\ell^\mrm{final}$, which we choose
to be a 1.85 arcmin FWHM Gaussian, and calculate the inverse spherical
harmonic transform of $\tilde{a}_{\ell m}^\mrm{SPT+\planck} M_{\ell m} B_\ell^\mrm{final}$.
This gives us the combined data maps. Note that the only smoothing and filtering
left in the combined maps is $B_{\ell}^\mrm{final}$, $M_{\ell m}$,
and the $\nside=8192$ pixel window function.


\section{Results} \label{sec:results}

In this section we present our main result: combined
data maps from SPT 95 GHz + \planck~100 GHz, SPT 150 GHz + \planck~143 GHz,
and SPT 220 GHz + \planck~217 GHz.
We also perform a few tests on the maps.

\subsection{Combined Data and Noise Maps}

\begin{figure*}[p]
\begin{centering}
\gridline{\fig{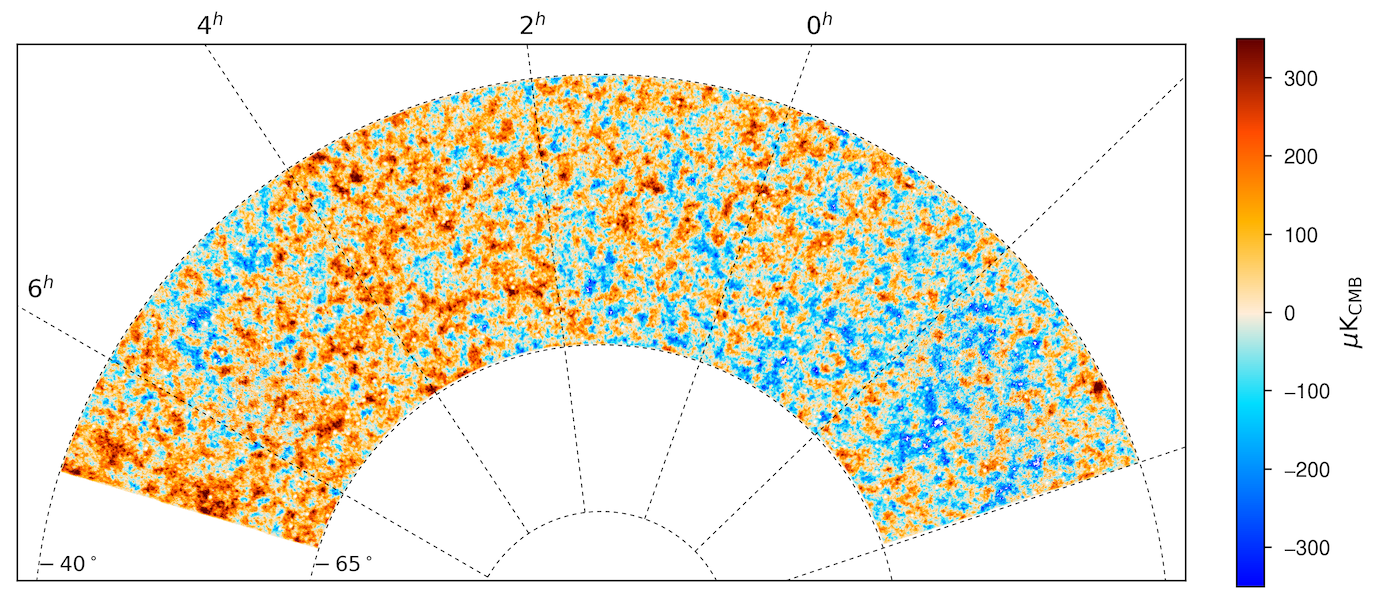}{\textwidth}{}
}
\gridline{\fig{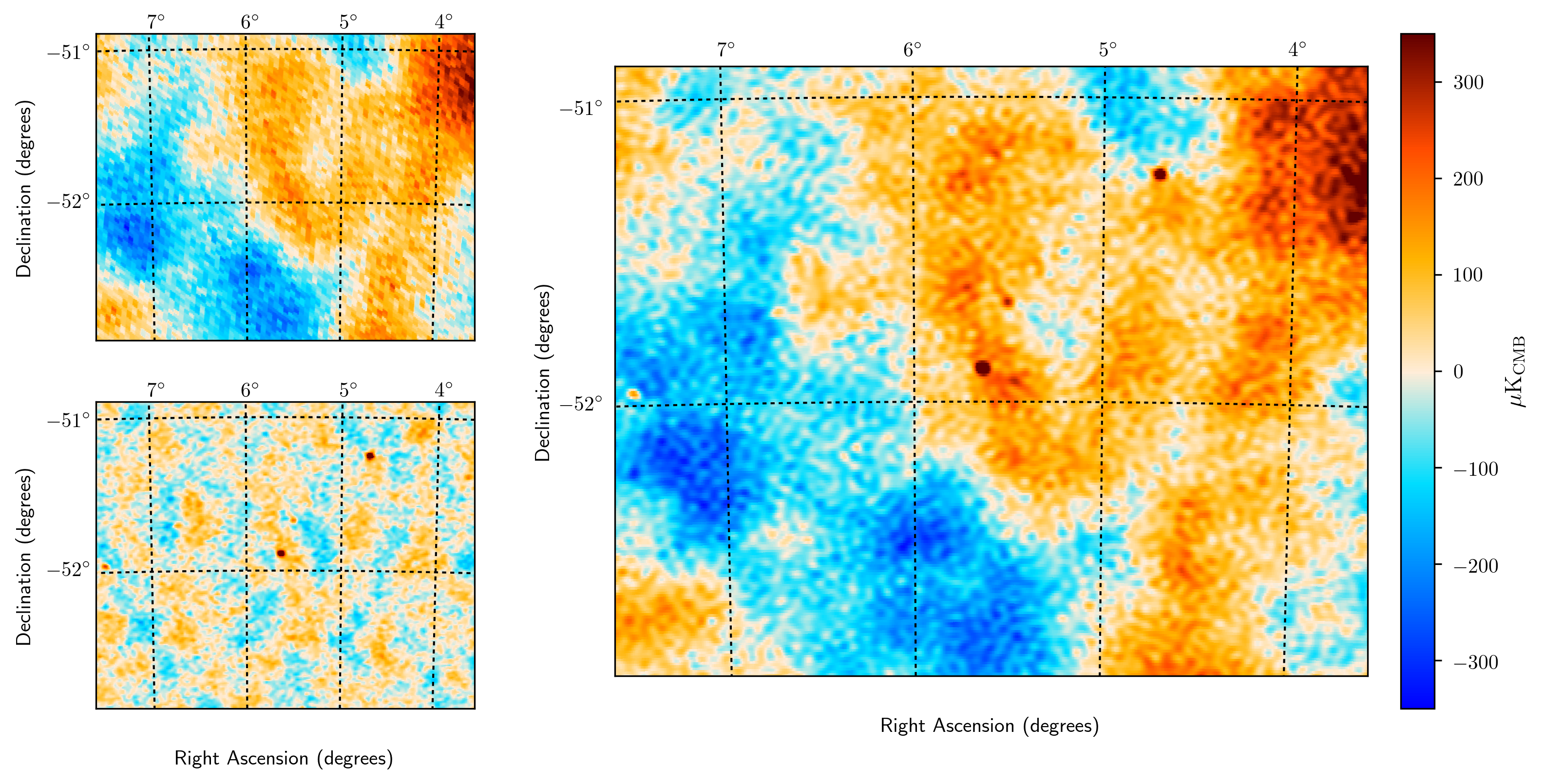}{\textwidth}{}
}
\caption{ Temperature map made from combined SPT-SZ 95 GHz and \planck~100 GHz data, plotted in the Lambert azimuthal equal-area projection (top), and zoomed in
on an $2^\circ \times 4^\circ$ patch at
(R.A., dec.) = (5.635$^\mrm{\circ}$, -51.883$^\mrm{\circ}$) (bottom).
The zoom-ins show SPT-only (far bottom left), \planck-only (left middle), and
SPT-\planck~(bottom right).
\label{fig:2500d_map90}}
\end{centering}
\end{figure*}

\begin{figure*}[p]
\begin{centering}
\gridline{\fig{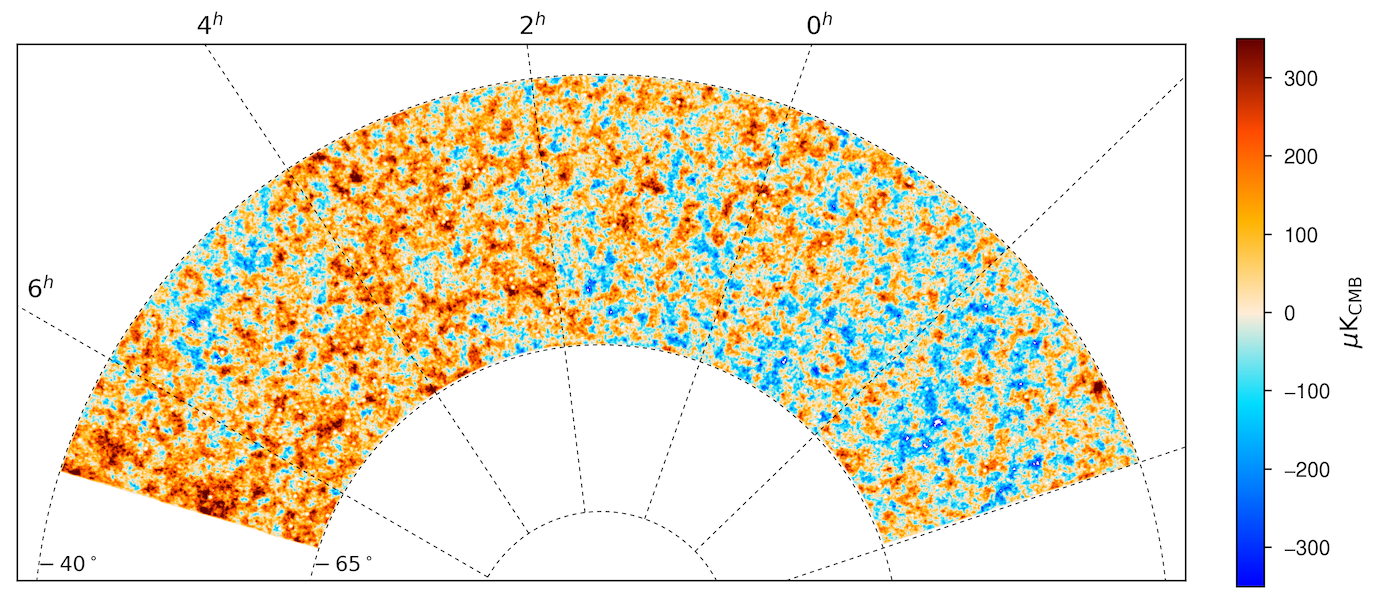}{\textwidth}{}
}
\gridline{\fig{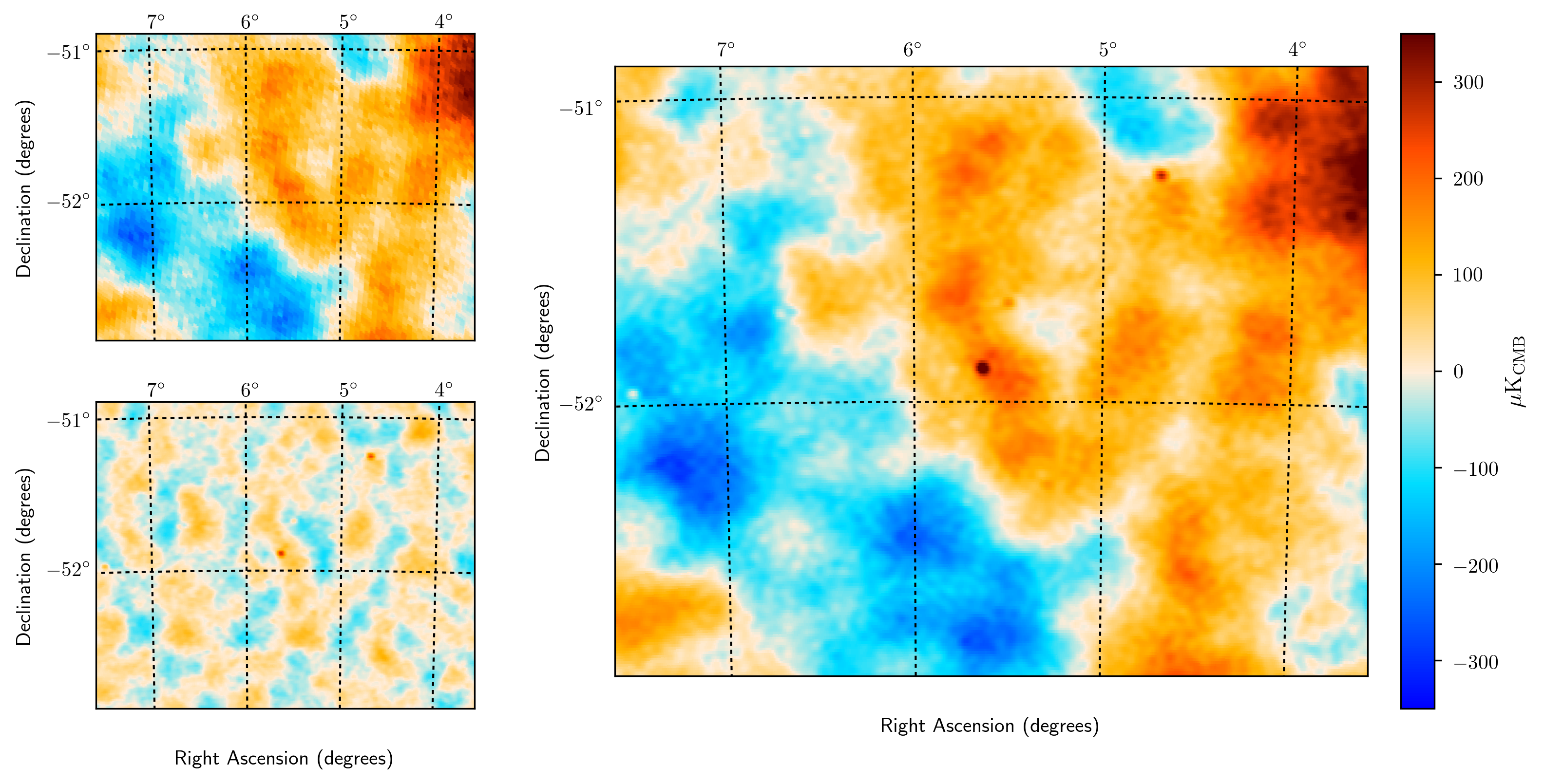}{\textwidth}{}
}
\caption{Temperature map made from combined SPT-SZ 150 GHz and \planck~143 GHz data, plotted in the Lambert azimuthal equal-area projection (top), and zoomed in
on an $2^\circ \times 4^\circ$ patch at
(R.A., dec.) = (5.635$^\mrm{\circ}$, -51.883$^\mrm{\circ}$) (bottom).
The zoom-ins show SPT-only (far bottom left), \planck-only (left middle), and
SPT-\planck~(bottom right).
\label{fig:2500d_map150}}
\end{centering}
\end{figure*}

\begin{figure*}[p]
\begin{centering}
\gridline{\fig{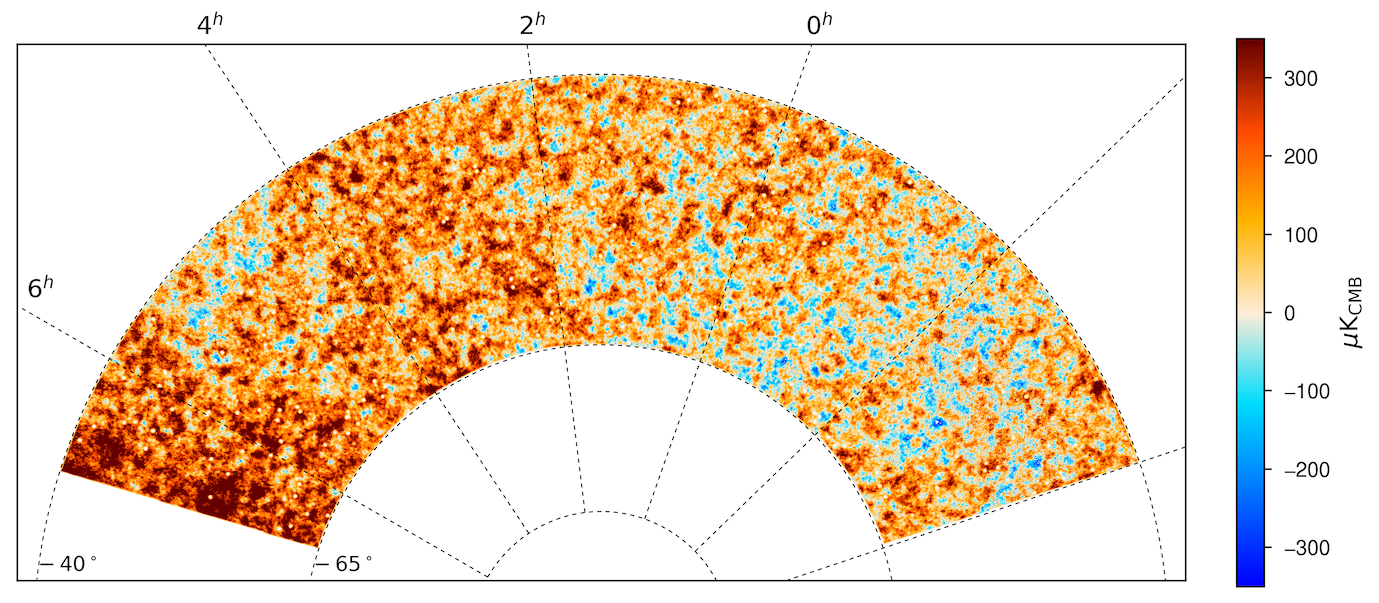}{\textwidth}{}
}
\gridline{\fig{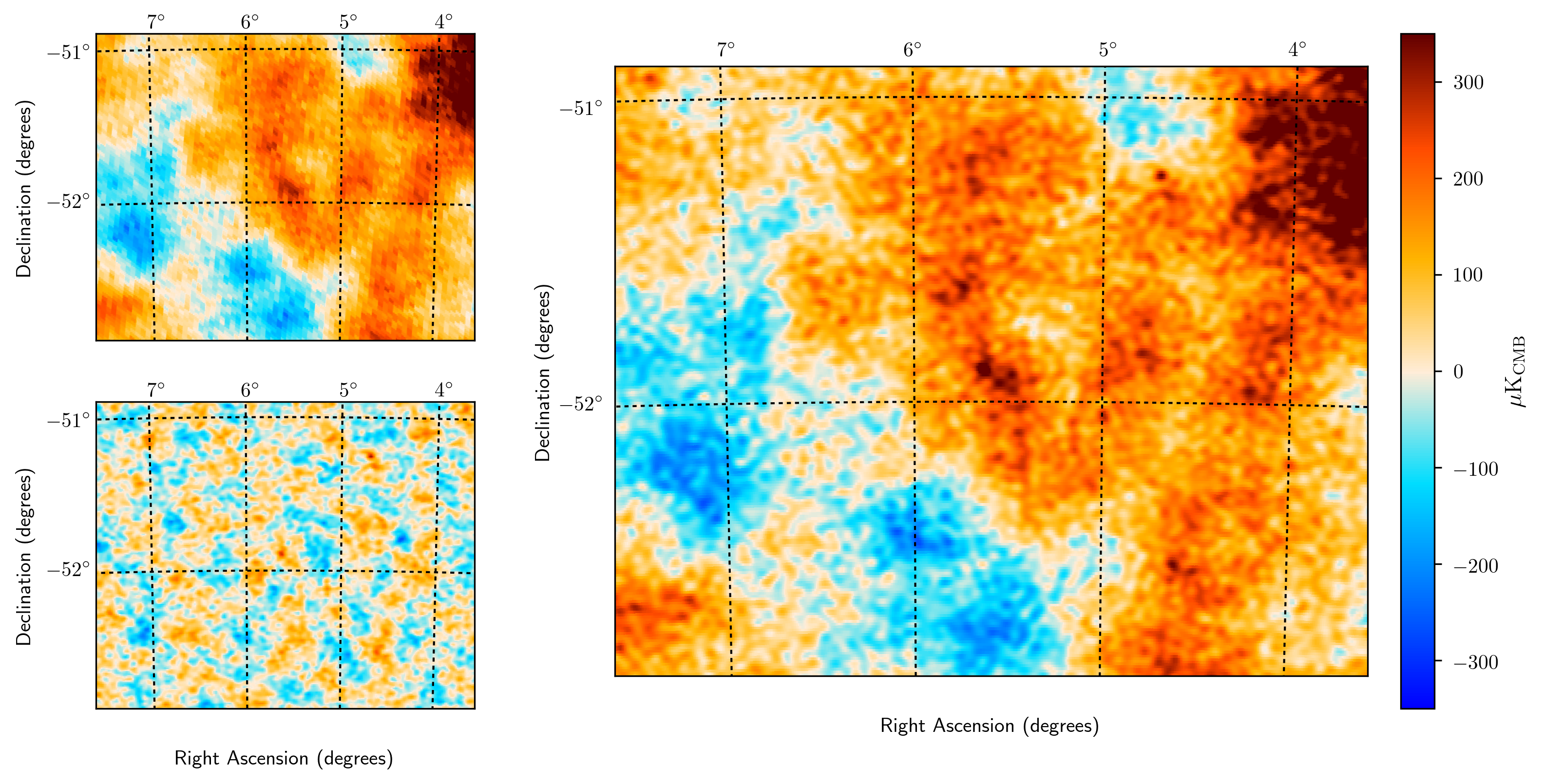}{\textwidth}{}
}
\caption{ Temperature map made from combined SPT-SZ 220 GHz and \planck~217 GHz data, plotted in the Lambert azimuthal equal-area projection (top), and zoomed in
on an $2^\circ \times 4^\circ$ patch at
(R.A., dec.) = (5.635$^\mrm{\circ}$, -51.883$^\mrm{\circ}$) (bottom).
The zoom-ins show SPT-only (far bottom left), \planck-only (left middle), and
SPT-\planck~(bottom right).
\label{fig:2500d_map220}}
\end{centering}
\end{figure*}

The combined data maps are shown in Figures~\ref{fig:2500d_map90},
\ref{fig:2500d_map150}, and~\ref{fig:2500d_map220}.
The degree-scale structure common to the three maps is primarily
the CMB; these large-scale features are contributed by \planck~data.
The diffuse emission which is brightest closer to the Galactic plane is
predominantly thermal dust in the Milky Way.
Small-scale foregrounds such as dusty star-forming galaxies and
radio sources, the finer-scale structure of the CMB, and the finer-scale structure
of Galactic emission, are all contributed mainly by SPT data.

We use the 100 noise simulations for SPT and \planck~to make
100 combined SPT-\planck~noise simulations for each band.
These are processed in the same way as the real combined data.
The average angular power spectrum of the noise for each band, 
which is computed as specified in Section~\ref{sec:powerspectra}, is plotted in Figure~\ref{fig:2500d_noise_1d}.

\subsection{Frequency Response of the Combined Maps}
 
\begin{figure*}[p]
\begin{centering}
\gridline{\fig{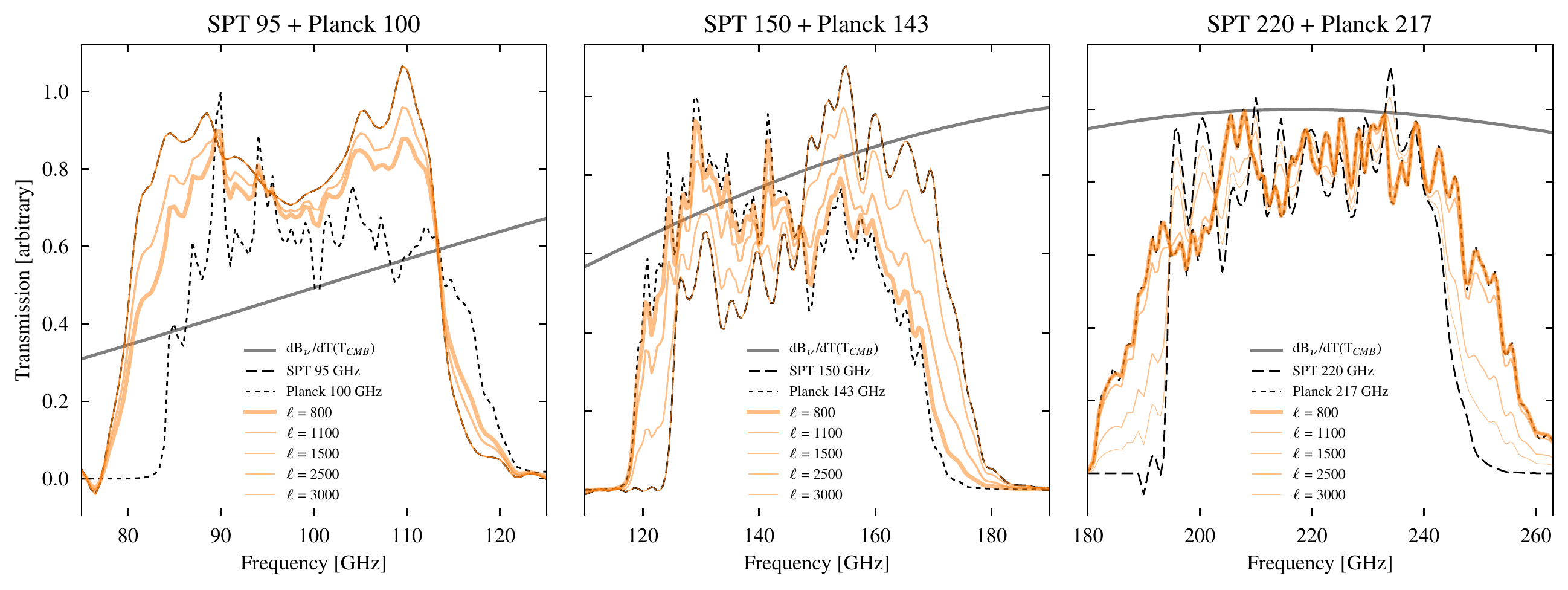}{\textwidth}{}
}
\caption{ Weighted combination of the SPT-SZ and \planck\ frequency response
functions at various values of $\ell$, such that the resulting curve represents
the frequency response in each combined map at each $\ell$.
The weights used are those at $m=600$. Note that at $m\lesssim 300$
the frequency response is given by the \planck\ frequency response, since 
SPT data is excluded at low values of $m$.
\label{fig:freqresp_combined}}
\end{centering}
\end{figure*}

As can be seen from Figure~\ref{fig:beams}b, the frequency response functions 
of the SPT-SZ and \planck\ data are slightly different. The frequency response
of the combined maps thus varies with $\ell$ and $m$, depending on the relative 
SPT-SZ and \planck\ weights.
Weighted bandpasses for selected values of $\ell$ (at $m=600$) 
are shown in Figure~\ref{fig:freqresp_combined}.
The most notable $\ell$ dependence in the bandpass functions are at the edges of the bands: 
the combined frequency response functions have slightly larger bandwidths than SPT 
or \planck\ alone. This is unlikely to substantially 
affect the interpretation of the signal in the combined maps, particularly in the low-foreground
SPT-SZ survey region.
The modes which are affected by this the most 
are where the weights are close to 0.5, which occurs where the noise power spectra
cross each other in Figure~\ref{fig:2500d_noise_1d}. At low and high $\ell$, 
the combined frequency response is essentially equal to that of 
\planck\ and SPT, respectively (except at low-$m$, where it is equal to that of \planck\ 
for all $\ell$).

\subsection{Tests}\label{sec:2500d_tests}

The upper halves of Figures~\ref{fig:2500d_map90},
\ref{fig:2500d_map150}, and~\ref{fig:2500d_map220} show the 2500 deg$^2$
SPT-\planck~combined temperature maps.
The gain in resolution achieved by combining SPT with \planck~data is
especially clear in the zoom-ins (the lower halves of Figures~\ref{fig:2500d_map90},
\ref{fig:2500d_map150} and~\ref{fig:2500d_map220}).
In the following subsections we present tests of the maps using
angular power spectra of simulated and real data.

\subsubsection{Tests Using Simulated Data}\label{sec:powerspectra}

As a test of our algorithm, we have compared the angular power spectrum of
simulated combined maps against the input power spectrum.
We mentioned in Section~\ref{sec:timestream} that $t_{\ell m}$ as defined
in Equation~\ref{eq:tlm} is an approximation to the true filtering. Now we test 
this approximation by measuring how accurately
we can recover the input power spectra from mock-observed maps after deconvolving
$t_{\ell m}$.
We also test that the combining algorithm does not bias the power spectrum.

The first part of this test consists of processing simulated noise-free maps in
the same way as the real data. Then, using the
``Spatially Inhomogeneous Correlation Estimator for Temperature and Polarization"
\citep[PolSpice;][]{chon2004}
code\footnote{See \url{http://www2.iap.fr/users/hivon/software/PolSpice/index.html} for the PolSpice code and documentation.} we compute
the angular power spectrum of each simulated map $C_\ell$,
which is usually plotted as $D_\ell \equiv \ell (\ell +1) C_\ell / 2\pi$.

We use the following procedure to
correct for the effect of $M_{\ell m}$ on measured power spectra.
Using a separate set of simulated maps where the true power spectrum is known ($C_\ell^\mrm{true}$),
we compute the estimated power spectrum of the simulated maps ($C_\ell^\mrm{est}$) after
applying the mask to the maps, computing the spherical harmonic transform,
applying the filter $M_{\ell m}$, and then inverting the harmonic transform.
We compute the ratio of the average biased spectrum to the true spectrum
\begin{equation}
m_\ell \equiv \frac{C_\ell^\mrm{true} }{\left\langle C_\ell^\mrm{est} \right\rangle},
\end{equation}
and multiply the estimated power spectrum of the combined maps (simulated
and real data) by $m_\ell$. We have checked using the separate set of simulated maps
that the uncertainty contributed by $m_\ell$ is negligible compared to other
sources of uncertainty.

\begin{figure*}
\begin{centering}
\gridline{\fig{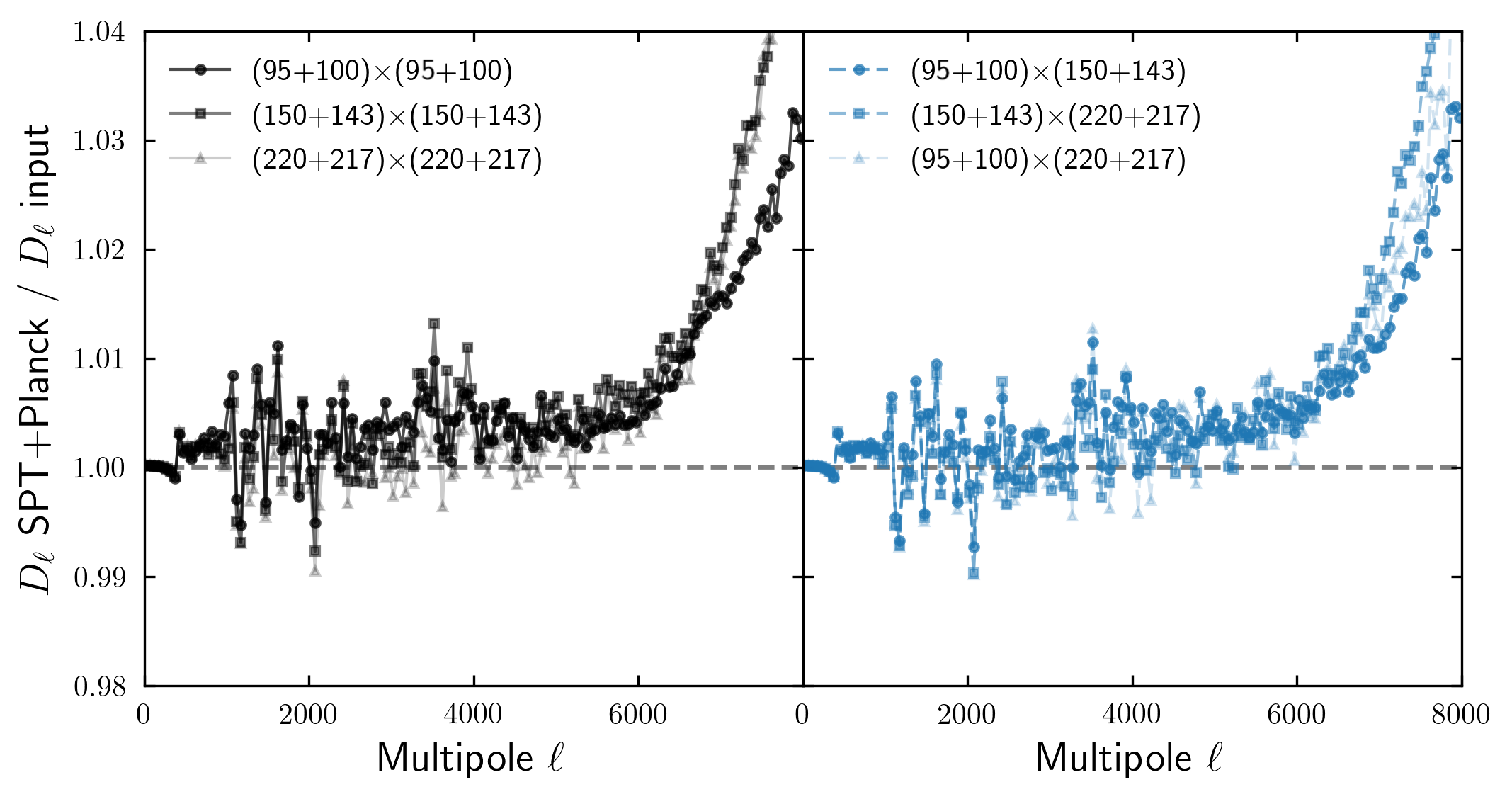}{0.8\textwidth}{}}
\caption{Angular power spectra of simulated noise-free SPT-\planck\ maps
divided by the average angular power spectra of the input maps.
The simulated data was processed identically to the real data.
On the left we show the auto-spectra relative to the
input auto-spectra, and on the right we show the cross-frequency spectra
relative to the average cross-frequency spectra of the input maps.
A ratio of 1.0 (dotted horizontal line) indicates perfect agreement between the spectra of the simulated maps
and the inputs.
All of the spectra were binned with bin widths of $\Delta \ell = 50$.
\label{fig:2500d_sptplanck_sim}}
\end{centering}
\end{figure*}

Using this procedure we compute auto- and cross-power spectra
averaged over a set of 100 simulations. The average power spectra are binned
and divided by their corresponding input spectra. These ratios are plotted in
Figure~\ref{fig:2500d_sptplanck_sim}. A ratio of $1.0$ indicates
perfect agreement between the spectra of simulated output and input.
For the majority of the $\ell$ range, especially where CMB is the dominant source
of fluctuations, the agreement is better than $1$ percent in power.
This test shows excess power in the simulations at high values of $\ell$ (most notably $\ell \gtrsim 6000$).
The excess power has similar $\ell$ dependence in auto spectra, cross spectra, and
across bands.
If our calculated transfer functions were noisy we would see excess power in the output simulations.
However, we found that the
excess power is unaffected by increasing the number of simulations that go into the transfer functions.
We also ran this test on each field separately and found that the
result does not change significantly from field to field. Together, these findings have led
us to believe that the departure from 1.0 at high $\ell$ is likely due to spatial variation in the
data weights. 
We recommend caution when using these maps above $\ell=6000$.

%

\subsubsection{Data Power Spectra}


We compare auto- and cross-spectra of our SPT-\planck\ data maps with
theoretical power spectra and previously published SPT-SZ power spectra.
Specifically, we compare to the power spectra published in S13, which
presented the 150~GHz auto-spectrum in the multipole range $650 < \ell < 3000$,
and in
\citet[][hereafter G15]{george15},
which presented auto and cross-spectra of SPT-SZ data in all three bands
in the multipole range $2000 < \ell < 11000$.
The theory spectra we compare to are the \planck\ 2015 best-fit CMB plus best-fit model
foreground spectra from G15.
These power spectrum comparisons are intended to validate the maps in
terms of the transfer functions and noise, not to estimate cosmological parameters.

We note that the target multipole range, and hence the
filtering of SPT-SZ data and masking of point sources, in this work
is matched to that in S13 and not G15, but we expect to be able to make
meaningful comparisons between the two analyses regardless.
Sources with 150 GHz flux densities greater than 6.4 mJy were masked
in the G15 mapmaking and power spectrum calculations. The maps presented
here were made with sources above 50 mJy at 150 GHz masked. For the comparison
to G15, we compute auto and cross power spectra using the same 6.4 mJy mask as in
that work.

\begin{figure*}
\gridline{\fig{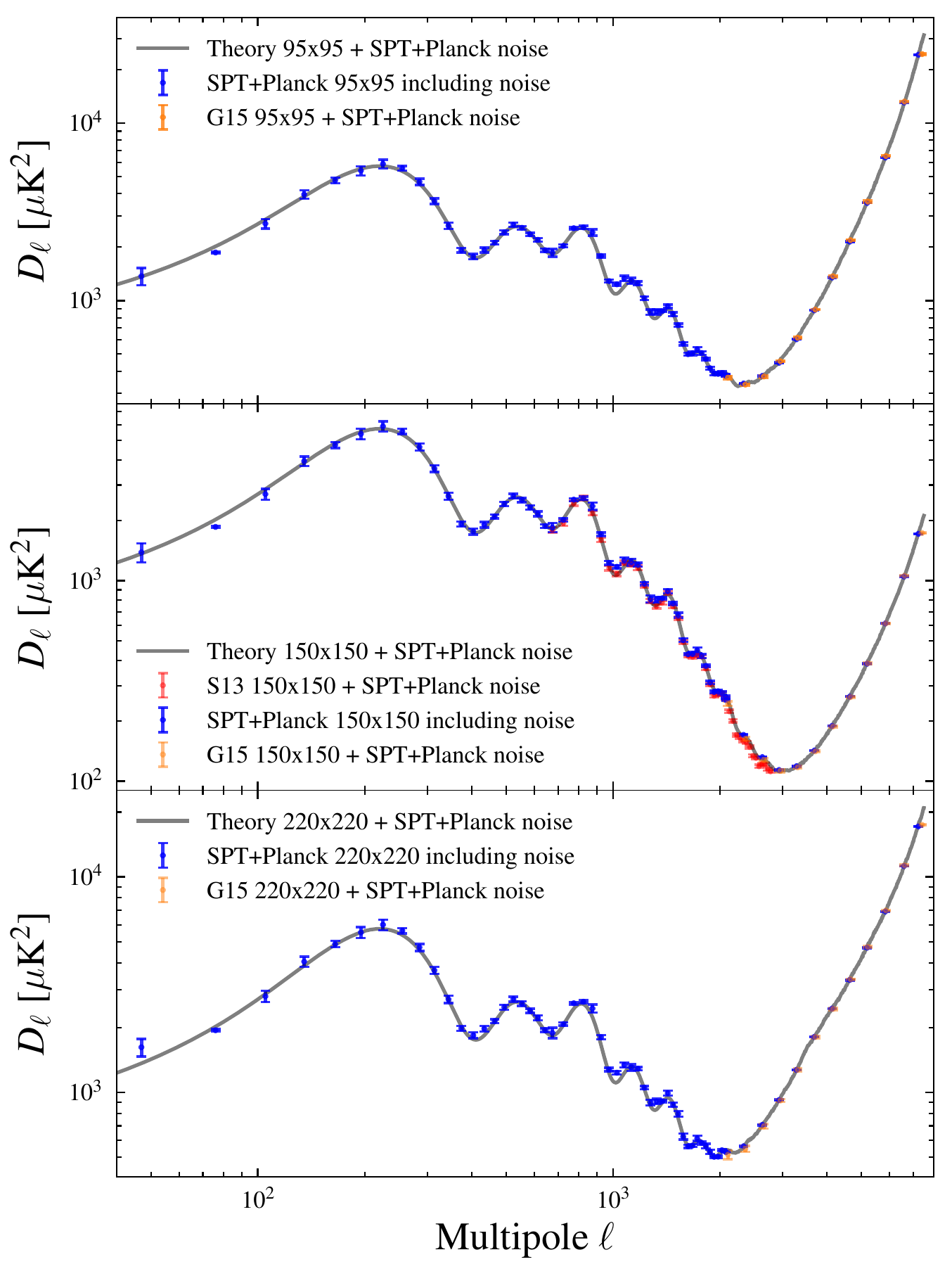}{0.75\textwidth}{}}
\caption{ Auto-spectra of the
combined SPT-\planck\ data maps using the 6.4 mJy mask, computed using PolSpice.
Theoretical auto-spectra (overlaid in grey) include our
best estimate of the statistics of the true sky (CMB + foregrounds).
The data auto-spectra are not noise-bias-subtracted. We have added the
noise power spectra to the theoretical auto-spectra, the G15 auto-spectra, and the S13 150 GHz
auto-spectrum.
The S13 auto-spectrum used a 50 mJy point source mask; we have subtracted
an estimate of the 6.4 mJy $\leq F_{150} \leq$ 50 mJy point source power from these
bandpowers.
\label{fig:data_autopowerspectra}}
\end{figure*}

\begin{figure*}
\gridline{\fig{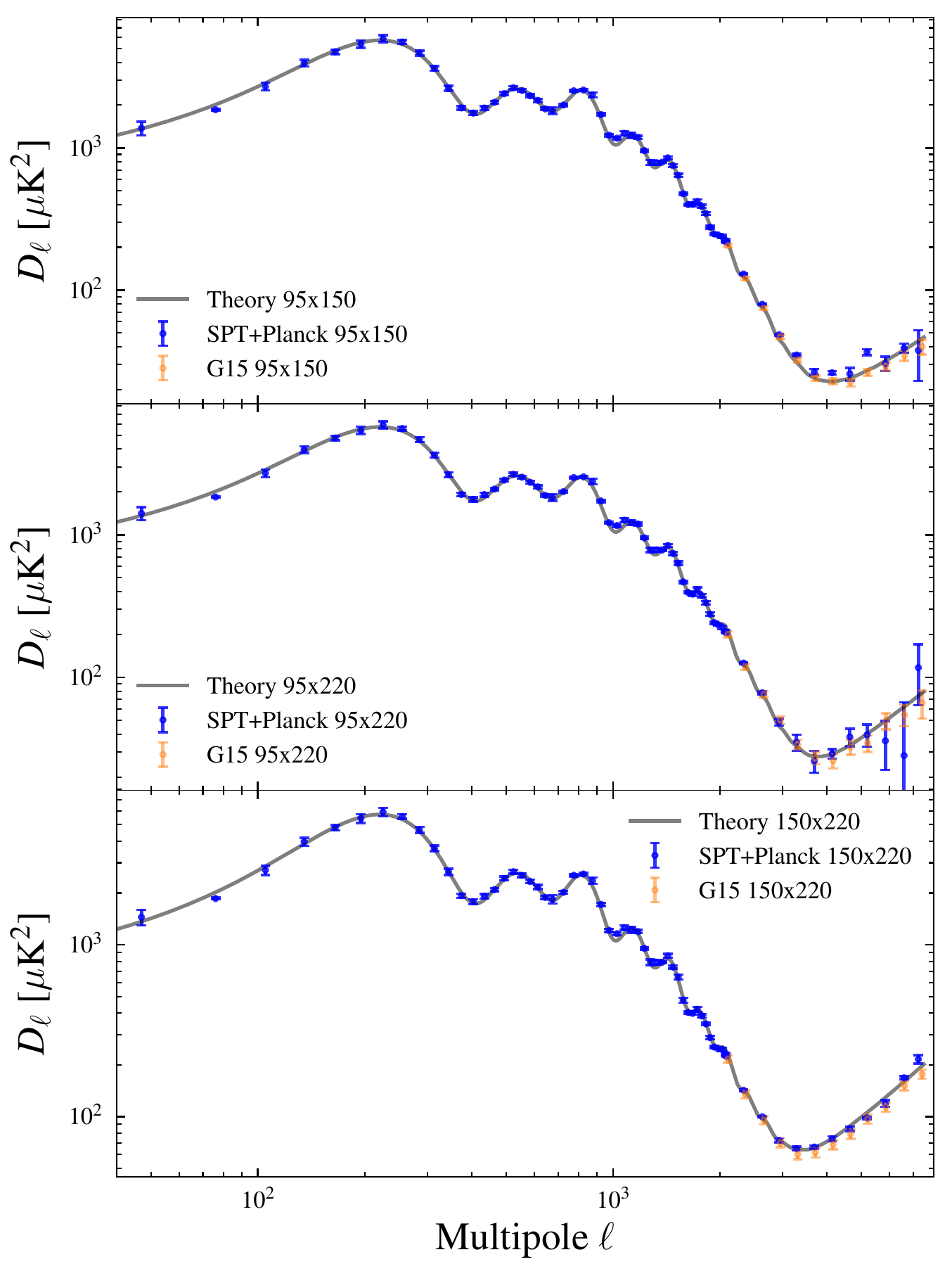}{0.75\textwidth}{}}
\caption{ Cross-spectra of the
combined SPT-\planck\ data maps using the 6.4 mJy mask, computed using PolSpice.
Theoretical cross-spectra (overlaid in grey) include our
best estimate of the statistics of the true sky (CMB + foregrounds),
including the cross-correlation of foregrounds between bands.
\label{fig:data_crosspowerspectra}}
\end{figure*}

The auto- and cross-spectra are shown in Figures~\ref{fig:data_autopowerspectra} and~\ref{fig:data_crosspowerspectra}, respectively.
The auto-spectra have not had noise bias subtracted; to compare to G15 and S13
spectra (which were computed using individual-observation cross-spectra and hence
do not suffer noise bias),
we have added the SPT-\planck\ noise power spectra to the published G15 and S13 auto-spectra
and to the theoretical spectra.
We have also applied a small correction to the S13 auto-spectrum to account for the
different point source mask used in S13.
The error bars are the standard deviation of the binned $D_\ell$ of noisy simulated maps.

We find that our cross-spectra are in good agreement with the theoretical
and G15 cross-spectra. The difference between our observed auto-spectra (signal plus noise)
and the expected spectra (theory plus noise spectra, or G15 auto-spectra plus noise spectra)
is found to be 3 \% of our noise power spectra. 



\section{Conclusions} \label{sec:conclusions}

We have made maps of the mm-wave southern sky from
combined SPT and \planck~data in three frequency bands.
The three final maps are created with a resolution described by a Gaussian
with FWHM of 1.85 arcmin, and individual sources measured to be 
brighter than 50 mJy at 150 GHz were masked from all three maps.
The 150 GHz SPT-SZ map was calibrated to the \planck\ 143 GHz data in the SPT-SZ patch 
\citep{hou2018}. The 95 GHz and 220 GHz SPT-SZ data were calibrated by inter-comparison with 
the 150 GHz data.
We determined the filter response functions of SPT-SZ data, 
and used previously measured SPT-SZ beam response functions to deconvolve the 
beam and filter response from 2500 deg$^2$ SPT-SZ maps in spherical harmonic space.
Estimates of the noise of each instrument were computed and used to combine 
the SPT and \planck\ data in spherical harmonic space. A small subset of modes
which are relatively noisy in \planck\ data and are not present in SPT data due to 
time stream filtering have been suppressed in the final maps so that the signal+noise power 
is approximately $\ell$-dependent.


The angular power spectra of simulated combined maps is found to agree very well
with input power spectra; this test shows a small amount of excess 
power (percent-level) in the simulated maps above $\ell \simeq6000$, so we recommend caution when
using the maps above this.
The auto and cross-power spectra of our combined data maps agree well with 
theoretical power spectra and previously published SPT-SZ power 
spectra \citep{story13, george15}. 

Along with the three combined data maps, we provide 5 realizations of the noise in
each combined map, and the mask that was applied to all the maps.
The maps and mask are in Equatorial coordinates, in \hpx\ format with $\nside=8192$ resolution.
We provide the two-dimensional filter that was applied to each combined map, as well as 
a Python script showing how one can apply a different filter to the maps if they wish.
All of the data products described in this paper are available at \url{https://pole.uchicago.edu/public/data/chown18/index.html}.

\acknowledgments The South Pole Telescope program is supported by the National Science Foundation through grant PLR-1248097. Partial support is also provided by the NSF Physics Frontier Center grant PHY-0114422 to the Kavli Institute of Cosmological Physics at the University of Chicago, the Kavli Foundation, and the Gordon and Betty Moore Foundation through Grant GBMF\#947 to the University of Chicago. CR acknowledges support from the Australian Research Council's Future Fellowships scheme (FT150100074). 
Work at Argonne National
Laboratory was supported under U.S. Department
of Energy contract DE-AC02-06CH11357.
This work has made use of computations performed on Guillimin, managed by
Calcul Quebec and Compute Canada (funded by CFI, MESI, and FRQNT).
The McGill authors acknowledge funding from the Natural Sciences and Engineering Research Council of Canada, Canadian Institute for Advanced Research, and Canada Research Chairs program.




\facilities{NSF/US Department of Energy 10m South Pole Telescope (SPT), 
European Space Agency (ESA) Planck Space Observatory Mission}

\software{\hpx/healpy \citep{gorski05}, PolSpice \citep{chon2004}, SciPy}

\bibliography{spt}



\end{document}